\documentclass{aa}
\usepackage{amssymb}
\usepackage{amsmath}
\usepackage{pdfsync}
\usepackage{txfonts}
\usepackage{balance}
\usepackage{natbib}
\bibpunct{(}{)}{;}{a}{}{,} % to follow the A&A style
\usepackage{color}
\usepackage{xspace}
\usepackage{bbm}
\usepackage{cancel}
\usepackage{graphicx}
\usepackage{ifthen}

\def\max{\mathrm{max}}

\def\rhogas{\ensuremath{\rho_\mathrm{g}}\xspace}
\def\rhodust{\ensuremath{\rho_\mathrm{d}}\xspace}

\def\mp{\ensuremath{m_\mathrm{p}}\xspace}

\newcommand{\e}[1]{\ensuremath{\times 10^{#1}}}
\newcommand{\Del}{\mathrm{d}}
\newcommand{\del}{\partial}
\newcommand{\ddel}[2]{\frac{\partial #1}{\partial #2}}
\newcommand{\half}{\frac{1}{2}}
\newcommand{\St}{\text{St}}
\newcommand{\rhos}{\rho_\text{s}}
\newcommand{\Sc}{\text{Sc}}
\newcommand{\uf}{\ensuremath{u_\text{f}}\xspace}
\newcommand{\cs}{\ensuremath{c_\mathrm{s}}\xspace}
\newcommand{\Hp}{\ensuremath{H_\mathrm{p}}\xspace}

\newcommand{\kb}{\ensuremath{k_\mathrm{b}}\xspace}
\newcommand{\Ok}{\ensuremath{\Omega_\mathrm{k}}\xspace}
\newcommand{\sigb}{\ensuremath{\sigma_\mathrm{B}}\xspace}
\newcommand{\tauros}{\ensuremath{\tau_\mathrm{R}}\xspace}
\newcommand{\taupla}{\ensuremath{\tau_\mathrm{P}}\xspace}
\newcommand{\kapros}{\ensuremath{\kappa_\mathrm{R}}\xspace}
\newcommand{\kappla}{\ensuremath{\kappa_\mathrm{P}}\xspace}
\newcommand{\Tm}{\ensuremath{T_\mathrm{mid}}\xspace}
\newcommand{\Siggas}{\ensuremath{\Sigma_\mathrm{g}}\xspace}
\newcommand{\Sigdust}{\ensuremath{\Sigma_\mathrm{d}}\xspace}
\newcommand{\nug}{\ensuremath{\nu_\mathrm{g}}\xspace}
\newcommand{\dx}{\ensuremath{\mathrm{d}}}
\newcommand{\N}{\ensuremath{\mathcal{N}}}
\makeatletter
\if@referee
    \newcommand{\com}[1]{\textcolor{red}{[{#1\xspace}]}}

    \usepackage[nolists,noheads]{endfloat}
\else
    \newcommand{\com}[1]{}

\fi
\makeatother
\usepackage[%
  pdfauthor={Tilman Birnstiel},
  pdftitle={Gas- and dust evolution in protoplanetary Disks},
  pdfstartview=FitH,
  linkcolor=blue,
  anchorcolor=blue,
  citecolor=blue,
  filecolor=blue,
  menucolor=blue,
  urlcolor=blue,
  colorlinks=true]{hyperref}
\defcitealias{Brauer:2008p215}{B08a}
\begin{document}
\title{Gas- and dust evolution in protoplanetary disks}
\titlerunning{Gas- and dust evolution in protoplanetary disks}
\author{T.~Birnstiel \and C.P.~Dullemond \and F.~Brauer\thanks{Deceased September 2009.}}
\authorrunning{T.~Birnstiel~et~al.}
\institute{Junior Research Group at the Max-Planck-Institut f\"ur Astronomie, K\"onigstuhl 17, D-69117 Heidelberg, Germany.\\Email: birnstiel@mpia.de } \date{\today}

\abstract
{Current models of the size- and radial evolution of dust in protoplanetary
disks generally oversimplify either the radial evolution of the disk (by focussing at one single radius or by using steady state disk models) or they assume particle growth to proceed monodispersely or without fragmentation. Further studies of protoplanetary disks -- such as observations, disk chemistry and structure calculations or planet population synthesis models -- depend on the distribution of dust as a function of grain size and radial position in the disk.}
{We attempt to improve upon current models to be able to investigate how the initial conditions, the build-up phase,  and the evolution of the protoplanetary disk influence growth and transport of dust.}
{We introduce a new version of the model of Brauer et al. (2008) in which we now include the time-dependent viscous evolution of the gas disk, and in which more advanced input physics and numerical integration methods are implemented.}
{We show that grain properties, the gas pressure gradient, and the amount of turbulence are much more influencing the evolution of dust than the initial conditions or the build-up phase of the protoplanetary disk. We quantify which conditions or environments are favorable for growth beyond the meter size barrier. High gas surface densities or zonal flows may help to overcome the problem of radial drift, however already a small amount of turbulence poses a much stronger obstacle for grain growth.}
{}


\maketitle

%\tableofcontents

% ==============================================================================
\section{Introduction}\label{sec:introduction}
The question of how planets form is one of the key questions in modern astronomy today. While it has been studied for centuries, the problem is still far from being solved. The agglomeration of small dust particles to larger ones is believed to be at least the first stage of planet formation. Both laboratory experiments \citep{Blum:2000p8110} and observations of the 10~$\mu$m spectral region \citep{Bouwman:2001p8118,vanBoekel:2003p8117} conclude that grain growth must take place in circumstellar disks. The growth from sub-micron size particles to many thousand kilometer sized planets covers 13 orders of magnitude in spatial scale and over 40 orders of magnitude in mass. To assemble a single 1 km diameter planetesimal requires the agglomeration of about $10^{27}$ dust particles. These dynamic ranges are so phenomenal that one has to resort to special methods to study the growth of particles though aggregation in the context of planet(esimal) formation.

A commonly used method for this purpose makes use of particle size distribution functions. The time dependent evolution of these particle size distribution functions has been studied by \citet{Weidenschilling:1980p4572}, \citet{Nakagawa:1981p4533}, \citet{Dullemond:2005p378}, \citet{Brauer:2008p215} (hereafter \citetalias{Brauer:2008p215}) and others. It was concluded that dust growth by coagulation can be very quick initially (in the order of thousand years to grow to centimeter sized aggregates at 1 AU in the solar nebula), but it stalls around decimeter to meter size due to what is known as the ``meter size barrier'': a size range within which particles achieve large enough velocities to undergo destructive collisions and fast radial inward drift toward the central star \citep{Weidenschilling:1977p865,Nakagawa:1986p2048}.

While the existence of this meter size barrier (ranging in fact from a couple of centimeters to tens of meters at 1 AU) has been known for over 30~years, a thorough study of this barrier, including all known mechanisms that induce motions of dust grains in protoplanetary disks, and at all regions in the disk, for various conditions in the disk and for different properties of the dust (such as sticking efficiency and critical fragmentation velocity), has been only undertaken recently \citepalias[see][]{Brauer:2008p215}. It was concluded that the barrier is indeed a very strong limiting factor in the formation of planetesimals from dust, and that special particle trapping mechanisms are likely necessary to overcome the barrier. 

However, this work was based on a static, non-evolving gas disk model. It is known that over the duration of the planet formation process the disk itself also evolves dramatically \citep{LyndenBell:1974p1945,Hartmann:1998p664,Hueso:2005p685}, which may influence the processes of dust coagulation and fragmentation. It is the goal of this paper to introduce a combined disk-evolution and dust-evolution model which also includes additional physics: we include relative azimuthal velocities, radial dependence of fragmentation critical velocities and the Stokes-drag regime for small Reynolds numbers.

The aim is to find out what the effect of disk formation and evolution is on the process of dust growth, how the initial conditions affect the final outcome, and whether certain observable signatures of the disk (for instance, its degree of dustiness at a given time) can tell us something about the physics of dust growth.

Moreover, this model will serve as a supporting model for complementary modeling efforts such as the modeling of radiative transfer in protoplanetary disks (which needs information about the dust properties for the opacities) and modeling of the chemistry in disks (which needs information about the total amount of dust surface area available for surface chemistry). In this paper we describe our model in quite some detail, and thus provide a basis for future work that will be based on this model.

Furthermore, additional physics, such photoevaporation or layered accretion can be easily included, which we aim to do in the near future.

As outlined above, this model includes already many processes which influence the evolution of the dust and the gas disk. However, there are several aspects we do not include such as back-reactions by the dust through opacity or collective effects \citet{Weidenschilling:1997p4593}, porosity effects \citep{Ormel:2007p7127}, grain charging \citep{Okuzumi:2009p7473} or the ``bouncing barrier'' (Zsom et al., in press).

This paper is organized as follows: Section~\ref{sec:model} will describe all components of the model including the radial evolution of gas and dust, as well as the temperature and vertical structure of the disk and the physics of grain growth and fragmentation. In Section~\ref{sec:results} we will compare the results of our simulations with previous steady-state disk simulations and review the aforementioned growth barriers. As an application, we show how different material properties inside and outside the snow line can cause a strong enhancement of dust within the snow line. Section~\ref{sec:discussion} summarizes our findings.
A detailed description of the numerical method along with results for selected test cases can be found in the Appendix.
% ==============================================================================
\section{Model}\label{sec:model}
The model presented in this work combines a 1D viscous gas disk evolution code and a dust evolution code, taking effects of radial drift, turbulent mixing, coagulation and fragmentation of the dust into account. 
We model the evolution of gas and dust in a vertically integrated way. The gas disk is viscously evolving after being built up by in-falling material from a collapsing molecular cloud core.

The radial distribution of grains is subject to gas drag, radial drift, and turbulent mixing. To which extend each effect contributes, depends on the grain/gas coupling of each grain size. By simultaneously modeling about 100--200 different grain sizes, we are able to follow the detailed evolution of the dust sub-disk being the superposition of all sizes of grain distributions.

So far, the evolution of the dust distribution depends on the evolving gas disk but not vice versa. A completely self consistently coupled code is a conceptually and numerically challenging task which will be the subject of future work.

% ------------------------------------------------------------------------------
\subsection{Evolution of gas surface density}\label{sec:gas}
Our description of the viscous evolution of the gas disk follows closely the models described by \citet{Nakamoto:1994p798} and \citet{Hueso:2005p685}. In this paper we shall therefore be relatively brief and put emphasis on differences between those models and ours.

The viscous evolution of the gas disk can be described by the continuity equation,
\begin{equation}
\frac{\del \Siggas }{\del t} - \frac{1}{r}\frac{\del}{\del r}\left( \Siggas \, r \, u_\mathrm{g}\right) = S_\mathrm{g},
\label{eq:ssd}
\end{equation}
where the gas radial velocity $u_\mathrm{g}$ is given by \citep[see][]{LyndenBell:1974p1945}
\begin{equation}
u_\mathrm{g} = - \frac{3}{\Siggas\sqrt{r}} \frac{\del}{\del r} \left( \Siggas \nu_\mathrm{g} \sqrt{r} \right).
\label{eq:u_gas}
\end{equation}
$\Siggas = \int_{-\infty}^{\infty} \rhogas(z) \dx z$ is the gas surface density, $r$ the radius along the disk mid-plane and \nug the gas viscosity. The source term on the right hand side of Eq.~\ref{eq:ssd}, denoted by $S_\mathrm{g}$ can be either infall of material onto the disk or outflow.

The molecular viscosity of the gas is too small to account for relevant accretion onto the star, the time scale of viscous evolution would be in the order of several billion years. 
Observed accretion rates and disk lifetimes can only be explained if turbulent viscosity drives the evolution of circumstellar disks. Therefore \citet{Shakura:1973p4854} parameterized the unknown viscosity as the product of a velocity scale and a length scale. The largest reasonable values for these scales in the disk are the pressure scale height $\Hp$
\begin{equation}
\Hp = \frac{\cs}{\Ok}
\end{equation}
and the sound speed $\cs$. Therefore the viscosity is written as
\begin{equation}
\nug = \alpha \: \cs \: \Hp,
\end{equation}
where $\alpha$ is the turbulence parameter and $\alpha \leq 1$.

Today it is generally believed that disks transport angular momentum by turbulence, however the source of this turbulence is still debated. The magneto-rotational instability is the most commonly accepted candidate for source of turbulence \citep{Balbus:1991p4932}. Values of $\alpha$ are expected to be in the range of $10^{-3}$ to some $10^{-2}$ \citetext{see \citealp{Johansen:2005p8425}; Dzyurkevich et al., in press}. Observations confirm this range with higher probability for larger values \citep[see][]{Andrews:2007p4967}.

If the disk becomes gravitationally unstable, large scale mechanisms of angular momentum transport such as through the formation of spiral arms come into play. The stability of the disk can be described in terms of the Toomre parameter \citep{Toomre:1964p1002}
\begin{equation}
Q = \frac{\cs \Ok}{\pi \,G \, \Siggas}.
\end{equation}
Values below a critical value of $Q_\text{cr} = 2$ describe a weakly unstable disk, which forms non-axisymmetric instabilities like spiral arms. $Q$ values below unity lead to fragmentation of the disk.

The effect of these non-axisymmetric structures is to transport angular momentum outward and rearranging the surface density in the disk so as to counteract the unstable configuration. This mechanism is therefore to a certain extent self-limiting. Values above $Q_\text{cr}$ are not influenced by instabilities, values below $Q_\text{cr}$ form instabilities which increase $Q$ until the disk is marginally stable again. Our model approximates this mechanism by increasing the turbulence parameter $\alpha$  according to the recipe of \cite{Armitage:2001p993},
\begin{equation}
\alpha(r) = \alpha_0 + 0.01  \left(  \left(\frac{Q_\text{cr}}{\min(Q(r),Q_\text{cr})}\right)^2 - 1  \right),
\label{eq:alpha_of_Q}
\end{equation}
where $\alpha_0$ is a free parameter of the model which is taken to be $10^{-3}$, unless otherwise noted.

During the time of infall onto the disk, we use a constant, high value of $\alpha = 0.1$ to mimic the effective redistribution of surface density during the infall phase which also increases the overall stability of the disk. Once the infall stops, we gradually decrease the turbulence parameter on a timescale of 10~000 years until it reaches its input value.

Eq. \ref{eq:ssd} is a diffusion equation, which is stiff. This means, one faces the problem that the numerical step of an explicit integration scheme goes $\propto \Delta r^2$ (where $\Delta r$ is the radial grid step size) which would make the simulation very slow. One possible solution to this problem is using the method of implicit integration. This scheme keeps the small time scales of diffusion i.e. the fast modes in check. We are not interested in these high frequency modes, but they would become unstable if we used a large time step. With an implicit integration scheme (see Section~\ref{app:alg_advdif}) the time step can be chosen larger without causing numerical instabilities, thus increasing the speed of the computation.

% ------------------------------------------------------------------------------
\subsection{Radial evolution of dust}\label{sec:dust}
If the average dust-to-gas ratio in protoplanetary disks is in the order of $10^{-2}$ (as found in the ISM), then the dust does not dynamically influence the gas, while the gas strongly affects the dynamics of the dust.

Thermally, however, the dust has potentially a massive influence on the gas disk evolution through its opacity, but we will not include this in this paper. Therefore the evolution of the gas disk can, in our approximation, be done without knowledge of the dust evolution, while the dust evolution itself {\em does} need knowledge of the gas evolution.

There might be regions, where dust accumulates (such as the mid-plane of the disk or dead-zones and snow-lines) and its influence becomes significant or even dominant but we will not include feedback of such dust enhancements onto the disk evolution in this paper.

For now, we want to focus on the equations of motion of dust particles under the assumption that gas is the dominant material by mass. The interplay between dust and gas can then be described in terms of a dimensionless coupling constant, the \textit{Stokes number} which is defined as
\begin{equation}
\St = \frac{\tau_\text{s}}{\tau_\text{ed}},
\label{eq:ST_general}
\end{equation}
where $\tau_\text{ed}$ is the eddy turn-over time and $\tau_\text{s}$ is the stopping time.

The stopping time of a particle is defined as the ratio of the momentum of a particle divided by the drag force acting on it. There are four different regimes for the drag force which determine the dust-to-gas coupling \citep[see][]{Whipple:1972p4621,Weidenschilling:1977p865}. Which regime applies to a certain particle, depends on the ratio between mean free path $\lambda_\text{mfp}$ of the gas molecules and the dust particle size $a$ (i.e. the Knudsen number) and also on the particle Reynolds-number $Re = {2 a u/\nu_\mathrm{mol}}$ with
$\nu_\mathrm{mol}$ being the gas molecular viscosity
\begin{equation}
\nu_\mathrm{mol} = \half \, \bar u \, \lambda_\text{mfp},
\end{equation}
$\bar u$ the mean thermal velocity. The mean free path is taken to be
\begin{equation}
\lambda_\text{mfp} = \frac{1}{n\:\sigma_{\mathrm{H}_2}}
\end{equation}
where $n$ denotes the mid-plane number density and $\sigma_{\mathrm{H}_2} =2 \e{-15}$~cm$^2$.

The different regimes\footnote{It should be noted that ``Stokes regime'' refers to the regime where the drag force on a particle is described by the Stokes law -- this is not directly related to the Stokes number.} are
\begin{equation}
\tau_\text{s} = 
\left\{
\begin{array}{lll}
\frac{\rho_\text{s}\: a}{\rho_\text{g}\:\bar u}&        \text{for }  \lambda_\text{mfp}/a\gtrsim\frac{4}{9} &       \text{Epstein regime}\\
\\
\frac{2 \rho_\text{s}\: a^2}{9 \nu_\mathrm{mol} \:\rho_\text{g}}&        \text{for } Re<1& \text{Stokes regime 1}\\
\\
\frac{2^{0.6}\:\rho_\text{s}\:a^{1.6}}{9 \nu_\mathrm{mol}^{0.6} \: \rho_\text{g}^{1.4}\:u^{0.4}} & \text{for } 1<Re<800& \text{Stokes regime 2}\\
\\
\frac{6 \rho_\text{s}\: a}{\rho_\text{g}\:u}&       \text{for } Re>800& \text{Stokes regime 3}\\
\end{array}\right.
\label{eq:stopping_time}
\end{equation}
Here $u$ denotes the velocity of the dust particle with respect to the gas, $\bar u = c_\text{s} \sqrt{{\pi}/{8}}$ denotes the mean thermal velocity of the gas molecules, $\rho_\text{s}$ is the solid density of the particles and $\rho_\text{g}$ is the local gas density.

For now, we will focus on the Epstein regime. To calculate the Stokes number, we need to know the eddy turn-over time. As noted before, our description of viscosity comes from a dimensional analysis. We use a characteristic length scale $L_\text{c}$ and a characteristic velocity scale $V_\text{c}$ of the eddies. This prescription is ambiguous in a sense that it does not specify if angular momentum is transported by large, slow moving eddies or by small, fast moving eddies, that is
\begin{equation}
\nug = (\alpha^{q} V_\text{c}) \cdot (\alpha^{1-q} L_\text{c}).
\end{equation}

This is rather irrelevant for the viscous evolution of the gas disk, since all values of $q$ lead to the same viscosity, but if we calculate the turn-over-eddy time, we get
\begin{equation}
\tau_\text{ed} = \frac{2\pi L_\text{c}}{V_\text{c}} = \alpha^{1-2q}\: \frac{1}{\Ok}.
\end{equation}
The Stokes number and therefore the dust-to-gas coupling as well as the
relative particle velocities strongly depend on the eddy turnover time and
therefore on $q$ . In this work $q$ is taken to be 0.5
\citep[following][]{Cuzzi:2001p2167,Schrapler:2004p2394} which leads to a
turn-over-eddy time of
\begin{equation}
\tau_\text{ed} = \frac{1}{\Ok}.
\end{equation}
The Stokes number then becomes
%\begin{equation}
%\St = \frac{\rho_\text{s} \: a}{\rho_\text{g} \bar u} \Ok = \frac{\rho_\text{s}\cdot a}{\rho_\text{g}\cdot H_\text{g}} \cdot \sqrt{\frac{\pi}{8}}.
%\end{equation}
%Using Eq. \ref{eq:rho_midplane} leads to
\begin{equation}
\St = \frac{\rho_\text{s}\cdot a}{\Siggas} \cdot \frac{\pi}{2} \qquad \text{for } a<\frac{9}{4}\lambda_\text{mfp}.
\label{eq:ST_epstein}
\end{equation}
The overall radial movement of dust surface density $\Sigdust$ can now be described by an advection-diffusion equation,
\begin{equation}
\ddel{\Sigdust}{t} + \frac{1}{r} \ddel{}{r} \Bigl( r \, F_\text{tot} \Bigr) = 0,
\end{equation}
where the total Flux, $F_\text{tot}$ has contributions from a diffusive and an advective flux.
The diffusive part comes from the fact that the gas is turbulent and the dust couples to the gas. The dust is therefore turbulently mixed by the gas. Mixing counteracts gradients in concentration, in this case it is the dust-to-gas ratio of each size that is being smoothed out by the turbulence. The diffusive flux can therefore be written as
\begin{equation}
F_\text{diff} = - D_\text{d} \: \ddel{}{r} \left(\frac{\Sigdust}{\Siggas}\right) \cdot \Siggas.
\end{equation}
The ratio of gas diffusivity $D_\text{g}$ over dust diffusivity $D_\text{d}$ is called the Schmidt number. We follow \cite{Youdin:2007p2021}, who derived
\begin{equation}
\Sc \equiv \frac{D_\text{g}}{D_\text{d}} = {1+\St^2},
\end{equation}
and assume the gas diffusivity to be equal to the turbulent gas viscosity \nug.

The second contribution to the dust flux is the advective flux,
\begin{equation}
F_\text{adv} = \Sigdust \cdot u_\text{r},
\end{equation}
where $u_r$ is the radial velocity of the dust. There are two contributions to it,
\begin{equation}
u_\text{r} = \frac{u_\text{g}}{1 + \St^2} - \frac{2 u_\text{n}}{\St + \St^{-1}}.
\label{eq:u_r_dust}
\end{equation}
The first term is a drag term which comes from the radial movement of the gas which moves with a radial velocity of $u_\mathrm{g}$, given by Eq.~\ref{eq:u_gas}. Since the dust is coupled to the gas to a certain extend, the radially moving gas is able to partially drag the dust along.

The second term in Eq. \ref{eq:u_r_dust} is the radial drift velocity with respect to the gas. The gas in a Keplerian disk does in fact move sub-keplerian, since it feels the force of its own pressure gradient which is usually pointing inwards. Larger dust grains do not feel this pressure and move on a keplerian orbit. Therefore, from a point of view of a (larger) dust particle, there exists a constant head wind, which causes the particle to loose angular momentum and to drift inwards. This depends on the coupling between the gas and the particle and is described by the second term in Eq. \ref{eq:u_r_dust}.
$u_n$ denotes the maximum drift velocity of a particle,
\begin{equation}
u_\text{n} = - E_\mathrm{d} \cdot \frac{ \ddel{P_g}{r}}{2 \: \rho_\text{g} \: \Ok},
\label{eq:u_eta}
\end{equation}
which has been derived by \citet{Weidenschilling:1977p865}. Here, we introduced a radial drift efficiency parameter $E_\mathrm{d}$. This parameter describes how efficient the radial drift actually is, as several mechanisms such as zonal or meridional flows might slow down radial drift. This will be investigated in Section~\ref{sec:drift_barrier}.

Putting all together, the time dependent equation for the surface density of one dust species of mass $m_i$ is given by
\begin{equation}
\ddel{\Sigdust^i}{t} + \frac{1}{r}\ddel{}{r}
\left\lbrace r \cdot \left[ \Sigdust^i \cdot u_r^i -  D^i_\text{d} \cdot \ddel{}{r} \left( \frac{\Sigdust^i}{\Siggas}\right) \cdot \Siggas \right] \right\rbrace = S_\mathrm{d}^i,
\label{eq:dustequation}
\end{equation}
where we have included a source term $S^i_\mathrm{d}$ on the right hand side which can be positive in the case of infall or re-condensation of grains and negative in the case of dust evaporation or outflows.
This source term does not include the sources caused by coagulation and fragmentation processes (see Section~\ref{sec:smolu}). All sources will be combined into one equation later which is implicitly integrated in an un-split scheme (see Section~\ref{app:alg_both}).

Note that both, the diffusion coefficient and the radial velocity depend on the Stokes number and therefore on the size of the particle.

\begin{figure}[thb]
\resizebox{\hsize}{!}{\includegraphics{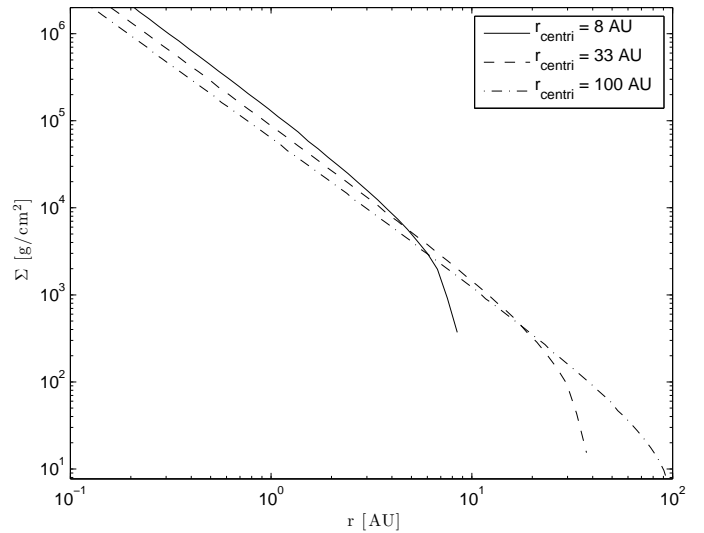}}
\caption{Total amount of in-fallen surface density as function of radius according to the Shu-Ulrich infall model (see Section~\ref{sec:collapse}) assuming a centrifugal radius of 8~AU (solid line), 33~AU (dashed line), and 100~AU (dash-dotted line).}
\label{fig:mass_loading}
\end{figure}

% ------------------------------------------------------------------------------
\subsection{Temperature and vertical structure}\label{sec:temperature}
The vertical structure can be assumed as being in hydrostatic equilibrium at all times if the disk is geometrically thin ($\Hp(r)/r\ll r$) and the vertical sound crossing time is much shorter than the radial drift time scale of the gas.
The isothermal vertical density structure is then given by
\begin{equation}
\rhogas(z) = \rho_0 \: \exp\left( - \half \: \frac{z^2}{\Hp^2} \right),
\end{equation}
where
\begin{equation}
\rho_0 = \frac{\Siggas}{\sqrt{2\:\pi} \: \Hp}.
\label{eq:rho_midplane}
\end{equation}
The viscous heating is given by \cite{Nakamoto:1994p798}
\begin{equation}
Q_+ = \Siggas \: \nug \left( r\: \ddel{\Ok}{r} \right)^2,
\end{equation}
were $\nug$ denotes the turbulent gas viscosity and $\Ok$ the Kepler frequency.

\citet{Nakamoto:1994p798} give a solution for the mid-plane temperature with an optically thick and an optical thin contribution,
\begin{equation}
{\Tm^4} = \frac{9}{8 \sigb} \left(   \frac{3}{8}{\tauros} + \frac{1}{2\: {\taupla}} \right)    \Siggas \: \alpha \: \cs^2\: \Ok + T_\text{irr}^4
\label{eq:t_mid}
\end{equation}
where we used $\nug = \alpha\, \cs\, \Hp$ and the approximation $\tau_\text{R/P} = \kappa_\text{R/P} \,\Siggas/2$. \kapros and \kappla are Rosseland and Planck mean opacities which will be discussed in the next section.

$T_\text{irr}$ contains contributions due to stellar or external irradiation.
Here, we use a formula derived by Ruden \& Pollack \citep[see][App. B]{Ruden:1991p1806},
\begin{equation}
T_\text{irr} =  T_\star \cdot \left[ \frac{2}{3 \pi} \left(\frac{R_\star}{r}\right)^3 + \half\: \left(\frac{R_\star}{r}\right)^2 \: \left(\frac{\Hp}{r}\right) \: \left(\frac{d\ln \Hp}{d\ln r} - 1 \right) \right]^{\frac{1}{4}},
\end{equation}
with a fixed $\mathrm{d ln}\Hp/\mathrm{d ln} r = 9/7$, following \citet{Hueso:2005p685}.
%Here, we use a simple approximation with constant flaring angle $\varphi$,
%\begin{equation}
%T_\text{irr} = T_\star \cdot  \left[ \half \varphi \left(\frac{R}{R_\star}\right)^2 \right]^\frac{1}{4}
%\end{equation}

The main source of opacity is the dust. Due to viscous heating, the temperature will increase with surface density. If the temperature rises above $\sim 1500$ K, the dust (i.e. the source of opacity) will evaporate, which stops the disk from further heating until all dust is vaporized. To simulate this behavior in our model, we calculate a gas temperature (assuming a small, constant value for gas opacity) in the case where the dust temperature rises above the evaporation temperature. Then $T_\text{mid}$ is approximated by
\begin{equation}
T_\text{mid} = \max(T_\text{gas},T_\text{evap.}),
\end{equation}
only if $T_\text{mid}$ from Eq.~\ref{eq:t_mid} would be larger than $T_\text{evap}$.

This is a thermostat effect: if $T$ rises above 1500 K, dust will evaporate, the opacity will drop and the temperature stabilizes at $T=1500$ K. Once even the very small gas opacity is enough to get temperatures $>1500$ K, all the dust is evaporated and the temperature rises further.

% ------------------------------------------------------------------------------
\subsection{Opacity}\label{sec:opacities}
In the calculation of the mid-plane temperature we use Rosseland and Planck mean opacities which are being calculated from a given frequency dependent opacity table. The results are stored in a look up table and interpolated during the calculations. The opacity table is for a mixture of 50\% silicates and 50\% carbonaceous grains.
%The mean opacities are defined as
%\begin{equation}
%\kapros(T) = \frac
%{\int\limits_0^\infty \frac{\del B_\nu(T)}{\del T}d\nu}
%{\int\limits_0^\infty\frac{1}{\kappa_\nu}\frac{\del B_\nu(T)}{\del T}d\nu},
%\end{equation}
%for the Rosseland mean opacity and
%\begin{equation}
%\kappla(T) = \frac
%{\int\limits_0^\infty\kappa_\nu B_\nu(T) d\nu}
%{\int\limits_0^\infty       B_\nu(T) d\nu},
%\end{equation}
%for the Planck mean opacity.

Since these are dust opacities, we convert them from \emph{cm$^2$/g dust} to \emph{cm$^2$/g gas} by multiplying the values with the dust-to-gas ratio $\epsilon_0$, which is a fixed parameter in our model, taken to be the canonical value of 0.01. This assumes that the mean opacity of the gas is very small and that the dust-to-gas ratio does not change with time. To calculate the opacity self-consistently, the total mass of dust and the distribution of grain sizes has to be taken into account, meaning that the dust evolution has a back reaction on the gas by determining the opacity. For now, our model does not take back-reactions from dust to gas evolution into account. Only in the case where the temperature rises above 1500 K, the drop of opacity due to dust evaporation is considered, as described above.

% ------------------------------------------------------------------------------
\subsection{Initial infall phase: cloud collapse}\label{sec:collapse}
The evolution of the protoplanetary disk also depends on the initial infall phase which builds up the disk from the collapse of a cold molecular cloud core. This process is still not well understood. First similarity solutions for a collapsing sphere were calculated by \cite{Larson:1969p2574} and \cite{Penston:1969p2601}. \cite{Shu:1977p843} found a self similar solution for a singular isothermal sphere. The Larson \& Penston solution predicts much larger infall rates compared to the inside-out collapse of Shu ($\dot m_\text{in} \approx 47\: c_\text{s}^3/G$ and $0.975\: c_\text{s}^3/G$ respectively).

More recent work has shown that the infall rates are not constant over time, but develop a peak of high accretion rates and drop to smaller accretion rates at later times. The maximum accretion rate is about $13\: c_\text{s}^3/G$ if opacity effects are included \citep[see][]{Larson:2003p3025}. Analytical, pressure-free collapse calculations of \cite{Myers:2005p4950} show similar behavior but with a smaller peak accretion rate of $\dot m_\text{in} = 7.07 \cs^3/G$. They also argue that the effects of pressure and magnetic fields will further increase the time scales of cloud collapse.

This initial infall phase is important since it provides the initial condition of the disk and also influences the whole simulation by providing a source of small grains and gas at larger distances to the star during later times of evolution.

It should be noted that several groups perform 3D hydrodynamic simulations of
collapsing cloud cores which show more complicated evolution
\citep[e.g.,][]{Banerjee:2006p8491,Whitehouse:2006p8532}. However, to be able to study general trends of the infall phase, we use the Shu-model since it is adjustable by a few parameters whose influences are easy to understand. In this model the collapse proceeds with an infall rate of $\dot m_\text{in}=0.975 \: c_\text{s}^3/G$ which stays constant throughout the collapse.

We assume the singular isothermal sphere of mass $M_\text{cloud}$ to be in solid body rotation $\Omega_\text{s}$. If in-falling material is conserving its angular momentum, all matter from a shell of radius $r_\text{s}$ will fall onto the star and disk system (with mass $m_\mathrm{cent}(t)$) within the so called centrifugal radius,
\begin{equation}
r_\text{centr}(t) = \frac{\Omega_\text{s}^2\: r_\text{s}^4}{G \: m_\text{cent}(t)},
\label{eq:r_centri}
\end{equation}
where $G$ is the gravitational constant and $r_\text{s} = 0.975\cdot c_\text{s}\:t /2$. The path of every parcel of gas can then be described by a ballistic orbit until it reaches the equatorial plane. The resulting flow onto the disk is
\begin{equation}
\dot\Sigma_\text{d}(r,t) = 2\;\rho_1(r,t) \cdot u_\text{z}(r,t),
\end{equation}
where
\begin{equation}
u_\text{z}(r,t) = \sqrt{\frac{G \: m_\text{cent}(t)}{r}} \cdot \mu,
\end{equation}
and
\begin{equation}
\rho_1(r,t) = \frac{\dot m_\mathrm{in}}{8 \pi \sqrt{G \: m_\text{cent}(t) \: r^3}} \cdot  \frac{r}{r_\text{centr}(r,t)} \cdot \frac{1}{\mu^2}
\end{equation}
as described in \cite{Ulrich:1976p856}.
Here, $\mu$ is given by
\begin{equation}
\mu = \sqrt{1-r/r_\text{centr}(r,t)}.
\end{equation}
The centrifugal radius can therefore be approximated by (cf. \citet{Hueso:2005p685})
\begin{equation}
  r_\mathrm{centri}(t) \simeq
  1.4 \left( \frac{\Omega_\mathrm{s}}{10^{-14} \mathrm{~s}^{-1}} \right)^2
  \left(\frac{m_\mathrm{cent}(t)}{M_\odot}\right)^3
  \left(\frac{\cs}{3\times 10^4 \mathrm{~cm~s}^{-1}}\right)^{-8}\mathrm{ AU}.
\end{equation}

We admit that this recipe for the formation of a protoplanetary disk is perhaps oversimplified. Firstly, as shown by \citet{Visser:2009p9087}, the geometrical thickness of the disk changes the radial distribution of in-falling matter onto the disk surface, because the finite thickness may ``capture'' an in-falling gas parcel before it could reach the midplane. Secondly, star formation is likely to be messier than our simple single-star axisymmetric infall model. And even in such a simplified scenario, the Shu inside-out collapse model is often criticized as being unrealistic. However, it would be far beyond the scope of this paper to include a better infall model. Here we just want to get a feeling for the effect of initial conditions on the dust growth, and we leave more detailed modeling to future work.

% ------------------------------------------------------------------------------
\subsection{Grain growth and fragmentation}\label{sec:growth_and_frag}
\begin{figure*}[t]
\centering
\resizebox{0.85\hsize}{!}{\includegraphics{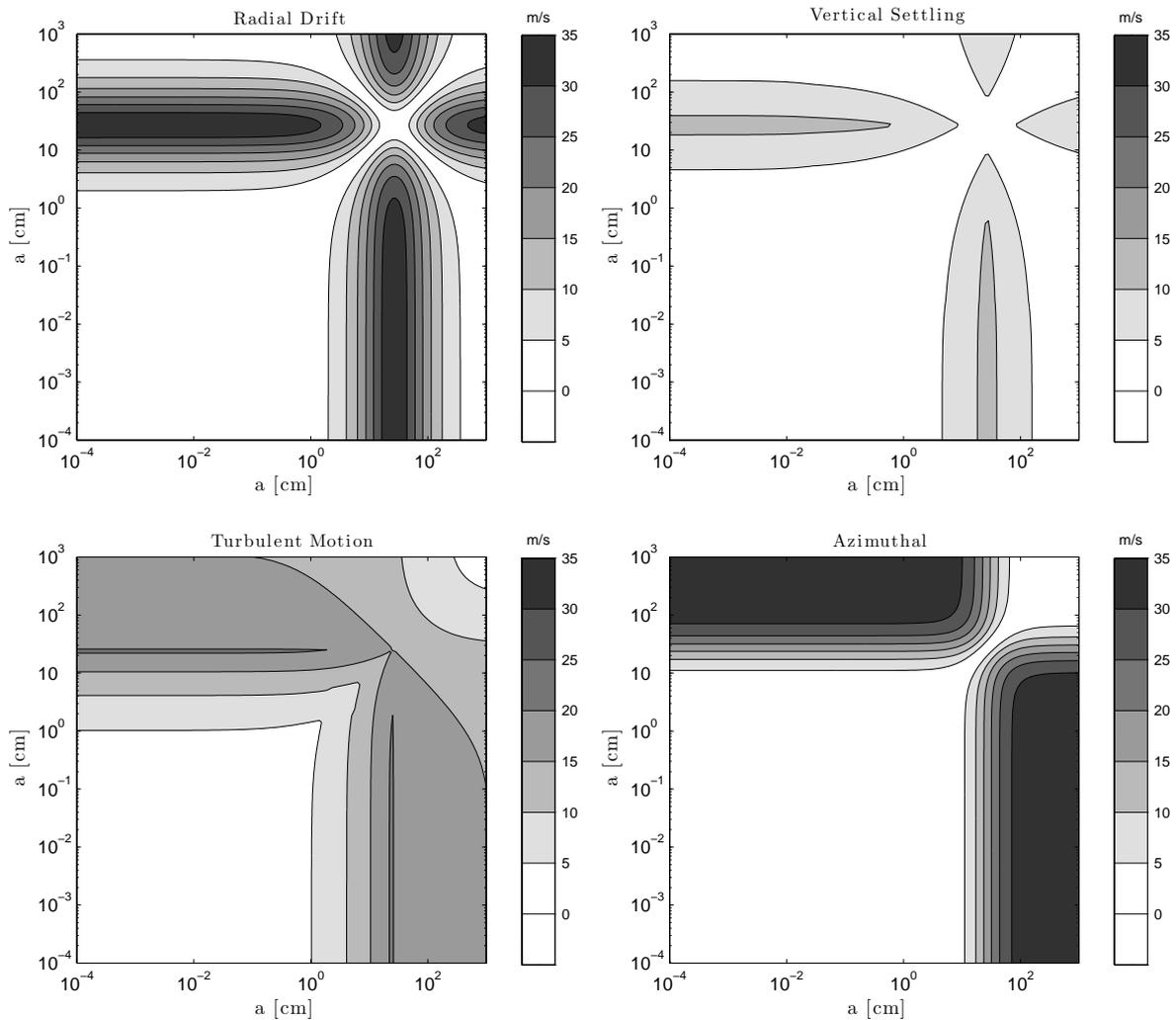}}
\caption{Sources of relative particle velocities considered in this model (Brownian motion velocities are not plotted) at a distance of 10~AU from the star. The turbulence parameter $\alpha$ in this simulation was $10^{-3}$. It should be noted that relative azimuthal velocities do not vanish for very large and very small particles.}
\label{fig:rel_vel}
\end{figure*}
The goal of the model described in this paper is to trace the evolution of gas and dust during the whole lifetime of a protoplanetary disk, including the grain growth, radial drift and turbulent mixing.

Here, the problem arises that radial drift and coagulation ``counteract'' each
other in the regime of $\St=1$ particles: coagulation of smaller sizes restores the population around $\St=1$, whereas radial drift preferentially removes these particles. To be able to properly model this behavior, the time step has to be chosen small enough if the method of operator splitting is used.

The upper limit for this time step can be very small. If larger steps are used the solution will ``flip-flop'' back and forth between the two splitted sub-steps of motion and coagulation, and the results become unreliable. A method to allow the choice of large time steps while preserving the accuracy is to use a non-splitted scheme in which the integration is done ``implicitly''. \citetalias{Brauer:2008p215} already use this technique to avoid flip-flopping between coagulation and fragmentation of grains, and we refer to that paper for a description of the general method. What is new in the current paper is that this implicit integration scheme is extended to also include the radial motion of the particles. So now radial motion, coagulation and fragmentation are done all within a single implicit integration scheme. See Appendix~\ref{app:alg_both} for details.

% ..............................................................................
\subsubsection{Smoluchowski equation}\label{sec:smolu}
The dust grain distribution $n(m,r,z)$, which is a function of mass $m$, distance to the star $r$ and height above the mid-plane $z$, describes the number of particles per cm$^3$ per gram interval in particle mass. This means that the total dust density in g cm$^{-3}$ is given by
\begin{equation}
\rho(r,z)=\int_0^\infty n(m,r,z) \cdot m \; \dx{m}.
\end{equation}

With this definition of $n(m,r,z)$, the coagulation/fragmentation at one point in the disk can be described by a general two-body process,
\begin{equation}
\begin{split}
\frac{\del}{\del t} n(m,r,z) =& \iint_0^\infty M(m,m',m'') \times\\
&\phantom{ \iint_0^\infty}\times n(m',r,z) \cdot n(m'',r,z) \,\dx m' \,\dx m'',
\end{split}
\label{eq:smolu}
\end{equation}
where $M(m,m',m'')$ is called the kernel. In the case of coagulation and fragmentation, this is given by
\begin{equation}
\begin{split}
M(m,m',m'')     =   &\phantom{+}\half K(m',m'') \cdot \delta(m'+m''-m)\\
                &- K(m',m'') \cdot \delta(m''-m)\\
                &+\half L(m',m'') \cdot S(m,m',m'')\\
                &- L(m',m'') \cdot \delta(m-m'').
\end{split}
\label{eq:combined_kernel}
\end{equation}
For better readability, the dependency of $M$ on radius and height above the mid-plane was omitted here. The combined coagulation/fragmentation kernel consists of four terms containing the coagulation kernel $K$, the fragmentation kernel $L$ and the distribution of fragments after a collision $S$.

The first two terms in Eq. \ref{eq:combined_kernel} correspond to gain (masses $m'$ and $m-m'$ coagulate) and loss ($m$ coagulates with $m'$) due to grain growth.

The third term describes the fragmentation of masses $m$ and $m'$, governed by the fragmentation kernel $L$ and the fourth term describes the fact that when masses $m'$ and $m''$ fragment, they distribute some of their mass via fragments to smaller sizes.

The coagulation and fragmentation kernels will be described in section
\ref{sec:coag_kernel}, the distribution of fragments, $S$, will be described in
the next section.

To be able to trace the size and radial evolution of dust in a combined way, we need to express all contributing processes in terms of the same quantity. Hence, we will formulate the coagulation/fragmentation equation in a vertically integrated way. The vertical integration can be done numerically \citepalias[as in][]{Brauer:2008p215}, however coagulation processes are most important near the mid-plane, which allows to approximate the kernels as being vertically constant (using the values at the mid-plane). If the vertical dust density distribution of each grain size is taken to be a Gaussian (see Section~\ref{sec:rel_velocities}, Eq.~\ref{eq:h_dust}), then the vertical integration can be done analytically, as discussed in Appendix~\ref{app:alg_coag}. Unlike the steady-state disk models of \citetalias{Brauer:2008p215} which have fixed surface density and temperature profiles, we need to recompute the coagulation and fragmentation kernels (which are functions of surface density and temperature) at every time step. Therefore this analytical integration also saves significant amounts of computational time.

We therefore define the vertically integrated dust surface density distribution
per logarithmic bin of grain radius, $\sigma(r,a)$ as
\begin{equation}
\sigma(r,a) = \int_{-\infty}^\infty n(r,z,a) \cdot m\cdot a\; \dx{z},
\label{eq:def_sigma}
\end{equation}
where $n(r,z,a)$ and $n(r,z,m)$ are related through $m=4\pi/3 \rhos a^3$.
The total dust surface density at any radius is then given by
\begin{equation}
\Sigma_\mathrm{d}(r) = \int_0^\infty \sigma(r,a)\; \mathrm{dln}a.
\end{equation}

Defining $\sigma(r,a)$ as in Eq.~\ref{eq:def_sigma} makes it a grid-independent density unlike the mass integrated over each numerical bin. This way, all plots of $\sigma(r,a)$ are meaningful without knowledge of the size grid that was used. Numerically, however we use the discretized values as defined in the appendix.

In our description of growth and fragmentation of grains, we always assume the
dust particles to have a constant volume density meaning that we do not trace the evolution of porosity of the particles as this is currently computationally too expensive with a statistical code as used in this work. This can be achieved with Monte-Carlo methods as in \citet{Ormel:2007p7127} or \citet{Zsom:2008p7126}, however these models have do not yet include the radial motion of dust and therefore cannot trace the global evolution of the dust disk.

% ..............................................................................
\subsubsection{Distribution of fragments}\label{sec:distr_of_fragments}
The distribution of fragments after a collision, $S(m,m',m'')$, is commonly described by a power law,
\begin{equation}
n(m) \text{d}m \propto m^{-\xi} \text{d}m.
\label{eq:frag_powerlaw}
\end{equation}

The value $\xi$ has been investigated both experimentally and theoretically.
Typical values have been found in the range between 1 and 2, by both
experimental \citep[e.g.,][]{Blum:1993p4324,Davis:1990p7995} and theoretical
studies \citep{Ormel:2009p8002}. Unless otherwise noted, we will follow \citetalias{Brauer:2008p215} by using the value of $\xi=1.83$.

In the case where masses of the colliding particles differ by orders of magnitude, a complete fragmentation of both particles is an unrealistic scenario. More likely, cratering will occur \citep{Sirono:2004p8225}, meaning that the smaller body will excavate a certain amount of mass from the larger one. The amount of mass removed from the larger one is parameterized in units of the smaller body $m_\text{s}$,
\begin{equation}
m_\text{out} = \chi \: m_\text{s}.
\end{equation}
The mass of the smaller particle plus the mass excavated from the larger one is then distributed to masses smaller than $m_\text{s}$ according to the distribution described by Eq. \ref{eq:frag_powerlaw}. In this work, we follow \citetalias{Brauer:2008p215} by assuming $\chi$ to be unity, unless otherwise noted.

% ..............................................................................
\subsubsection{Coagulation and fragmentation kernels}\label{sec:coag_kernel}
The coagulation kernel $K(m_1,m_2)$ can be factorized into three parts,
\begin{equation}
K(m_1,m_2) = \Delta u(m_1,m_2) \: \sigma_\mathrm{geo}(m_1,m_2) \: p_\text{c},
\end{equation}
and, similarly, the fragmentation kernel can be written as
\begin{equation}
L(m_1,m_2) = \Delta u(m_1,m_2) \: \sigma_\mathrm{geo}(m_1,m_2) \: p_\text{f}.
\end{equation}
Here, $\Delta u(m_1,m_2)$ denotes the relative velocity of the two particles, $\sigma_\mathrm{geo}(m_1,m_2)$ is the geometrical cross section of the collision and $p_\text{c}$ and $p_\text{f}$ are the coagulation and fragmentation probabilities which sum up to unity. In general, all these factors can also depend on other material properties such as porosity, however we always assume the dust grains to have a volume density of $\rhos=1.6$~g~cm$^{-3}$.

The fragmentation probability is still not well known and subject of both theoretical \citep{Paszun:2009p8871,Wada:2008p4903} and experimental research \citep[see][]{Blum:2008p1920,Guttler:2009p8384}. In this work, we adopt the simple recipe
\begin{equation}
p_\text{f} = \left\{
\begin{array}{ll}
0&                              \text{if } \Delta u < \uf - \delta u\\
\\
1&                              \text{if } \Delta u > \uf\\
\\
1-\frac{\uf-\Delta u}{\delta u}&    \text{else}
\end{array}
\right.
\end{equation}
with a transition width $\delta u$ and the fragmentation speed \uf as free parameter which is assumed to be 1 m~s$^{-1}$, following experimental work of \citet{Blum:1993p4324} and theoretical studies of \citet{Leinhardt:2009p5282}.

% ..............................................................................
\subsubsection{Relative particle velocities}\label{sec:rel_velocities}
The different sources of relative velocities considered here are Brownian motion, relative radial and azimuthal velocities, turbulent relative velocities and differential settling. These contributions will be described in the following, an example of the most important contributions is shown in Figure~\ref{fig:rel_vel}.

\textit{Brownian motion}, the thermal movement of particles, dependents on the mass of the particle. Hence, particles of different mass have an average velocity relative to each other which is given by
\begin{equation}
\Delta u_\text{BM} = \sqrt{\frac{8 k_\text{B} \: T (m_1 + m_2)}{\pi \: m_1 \: m_2 }}.
\end{equation}
Particle growth due to Brownian relative motion is most effective for small particles.

\textit{Radial drift}, as described in section \ref{sec:dust} also induces relative velocities since particles of different size are differently coupled to the gas. The relative velocity is then
\begin{equation}
\Delta u_\text{RD} = \left| u_\text{r}(m_1) - u_\text{r}(m_2) \right|,
\end{equation}
where the radial velocity of the dust, $u_\text{r}$ is given by Eq. \ref{eq:u_r_dust}.

\textit{Azimuthal relative velocities} are induced by gas drag in a similar way as radial drift. However while only particles (plus/minus 2 orders of magnitude) around \St=1 are significantly drifting, relative azimuthal velocities do not vanish for encounters between very large and vary small particles (see Figure~\ref{fig:rel_vel}). Consequently, large particles are constantly suffering high velocity impacts of smaller ones. According to \citet{Weidenschilling:1977p865} and \citet{Nakagawa:1986p2048}, the relative azimuthal velocities for gas-dominated drag are
\begin{equation}
 u_\varphi = \left| u_\mathrm{n} \cdot \left( \frac{1}{1+\St_1^2} - \frac{1}{1+\St_2^2} \right) \right|,
 \label{eq:dv_az}
\end{equation}
where $u_\mathrm{n}$ is defined by Eq.~\ref{eq:u_eta}.

\textit{Turbulent motion} as source of relative velocities is discussed in detail in \cite{Ormel:2007p801}. They also derive closed form expressions for the turbulent relative velocities which we use in this work.

\textit{Differential settling} is the fifth process we consider contributing to relative particle velocities. \cite{Dullemond:2004p390} constructed detailed models of vertical disk structure describing the depletion of grains in the upper layers of the disk. They show that the equilibrium settling speed for particles in the Epstein regime is given by
$u_\text{sett} = - z \: \Ok \: \St$ which can be derived by equating the frictional force $F_\text{fric} = - m \: u / t_\text{fric}$ and the vertical component of the gravity force, $F_\text{grav} = - m \: z \: \Ok^2$.
To limit the settling speed to velocities smaller than half the vertically projected Kepler velocity, we use
\begin{equation}
u_\text{sett} = - z \: \Ok\:\min\left(\St,\half\right)
\label{eq:u_settling}
\end{equation}
for calculating the relative velocities.

Since we do not resolve the detailed vertical distribution of particles, we take the scale height of each dust size as average height above the mid-plane, which gives
\begin{equation}
\Delta u_\text{DS} = \left| h_i \cdot \min(\St_i,1/2)  -  h_j  \cdot \min(\St_j,1/2)  \right| \: \cdot \Ok.
\end{equation}

\begin{figure}[thb]
\centering
\resizebox{0.8\hsize}{!}{\includegraphics{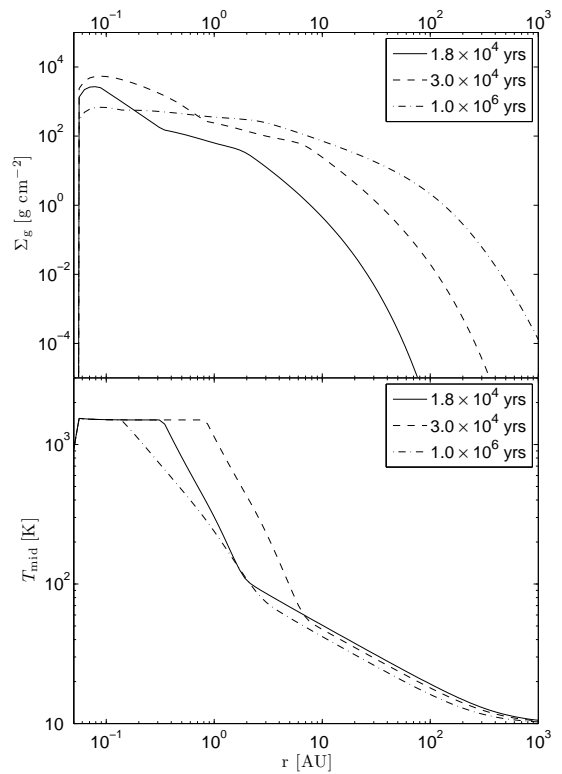}}
\caption{Evolution of disk surface density distribution (top) and midplane temperature (bottom) of the fiducial model described in
\ref{sec:visc_evol_of_gas_disk}.}
\label{fig:gas_T_snapshots}
\end{figure}

The dust scale height is calculated by equating the time scale for settling,
\begin{equation}
t_\text{sett} = \frac{z}{u_\text{sett}}
\end{equation}
and the time scale for stirring \citep{Dubrulle:1995p300,Schrapler:2004p2394,Dullemond:2004p390},
\begin{equation}
t_\text{stir} = \frac{z^2}{D_\text{d}}.
\end{equation}

By limiting the vertical settling velocity as in Eq. \ref{eq:u_settling} and by constraining the dust scale height to be at most equal to the gas scale height, one can derive the dust scale height to be
\begin{equation}
h_\text{d} = \Hp \cdot \min\left(1, \sqrt{\frac{\alpha}{\min(\St,1/2)(1+\St^2)}} \right).
\label{eq:h_dust}
\end{equation}

This prescription is only accurate for the dust close to the mid-plane, however most of the dust (and hence most of the coagulation/fragmentation processes) take place near the mid-plane, therefore this approximation is accurate enough for our purposes.

% ..............................................................................
%\subsubsection{Evaporation and re-condensation of dust}\label{sec:evaporation}
%At distances close enough to the star, temperatures are too high for dust to exist, particles drifting into this zone will be evaporated. The vaporized dust is not subject to drift and will therefore partly be mixed back out of the evaporation zone where it can re-condense and contribute to grain growth again. To include this mechanism in a simplified way, we assume that dust grains cannot exist at temperatures higher than 1500~K and that the timescale on which particles evaporate is independent on temperature, pressure and grain size.
%Every grain size is assumed to `decay' to the smallest grain size (which is small enough to behave like vapor) with a half-life of $\tau_\text{evap}$. The dust evaporation can then be described by
%\begin{equation}
%\frac{\del \Sigma_i}{\del t} = \left\{
%\begin{array}{ll}
%\displaystyle \sum_{i=2}^N \frac{\Sigma_i}{\tau_\text{evap}}&\quad\text{for $i=1$}\\
%\\
%\displaystyle-\frac{\Sigma_i}{\tau_\text{evap}}&\quad\text{otherwise}\\
%\end{array}
%\right.
%\end{equation}

% ==============================================================================
\section{Results}\label{sec:results}

% ------------------------------------------------------------------------------
\subsection{Viscous evolution of the gas disk}\label{sec:visc_evol_of_gas_disk}
We will now focus on the evolution of a disk around a T Tauri like star.
We assume the rotation rate of the parent cloud core to be $7\e{-14}$~s$^{-1}$,
which corresponds to 0.06 times the break up rotation rate of the core.
The disk is being built-up from inside out due to the Shu-Ulrich
infall model, with the centrifugal radius being 8~AU.
The parameters of our fiducial model are summarized in
Table~\ref{tab:model_parameters}.

\begin{figure}[thb]
\resizebox{\hsize}{!}{\includegraphics{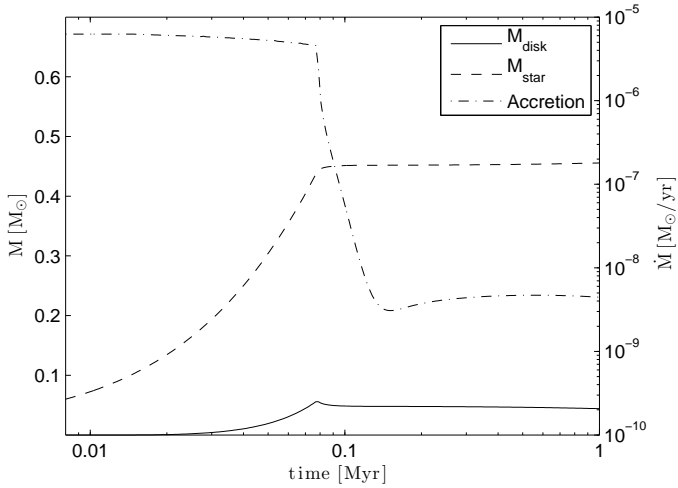}}
\caption{Evolution of disk mass and stellar mass (solid and dashed line on left scale respectively) and accretion rate onto the star (dash-dotted line on right scale). Adapted from Figure~5 in \citet{Hueso:2005p685}.}
\label{fig:accretion}
\end{figure}

Figure~\ref{fig:gas_T_snapshots} shows how the gas surface
density and the mid-plane temperature of this model evolve as the disk gets
built up, viscously spreads and accretes onto the star. It can be seen that
viscous heating leads to a strong increase of temperature at small radii. This
effect becomes stronger as the disk surface density increases during the infall
phase. After the infall has ceased, the surface density and therefore also the
amount of viscous heating falls off.

This effect also influences the position of the dust
evaporation radius, which is assumed to be at the radius where the dust
temperature exceeds 1500~K. This position moves outwards during the infall
(because of the stronger viscous heating described above). Once the infall
stops, the evaporation radius moves back to smaller radii as the large surface
densities are being accreted onto the star.

Figure~\ref{fig:accretion} shows the evolution of accretion rate onto the star, stellar mass
and disk mass. The infall phase lasts until
about 150\,000~years. At this point, the disk looses its source of gas and
small-grained dust and the disk mass drops off quickly until the disk has 
adjusted itself to the new condition. This also explains the sharp drop of the
accretion rate. The slight increase in the accretion rate afterwards comes from
the change in $\alpha$ after the infall stops (see Section~\ref{sec:gas}).
\citet{Hueso:2005p685} find a steeper, power-law decline of the accretion rate
after the end of the infall phase because their model does not take the effects
of gravitational instabilities into account.

% ------------------------------------------------------------------------------
\subsection{Fiducial model without fragmentation}\label{sec:fiducial_model_nofrag}

\begin{figure*}[htp]
\centering
\resizebox{0.9\hsize}{!}{\includegraphics{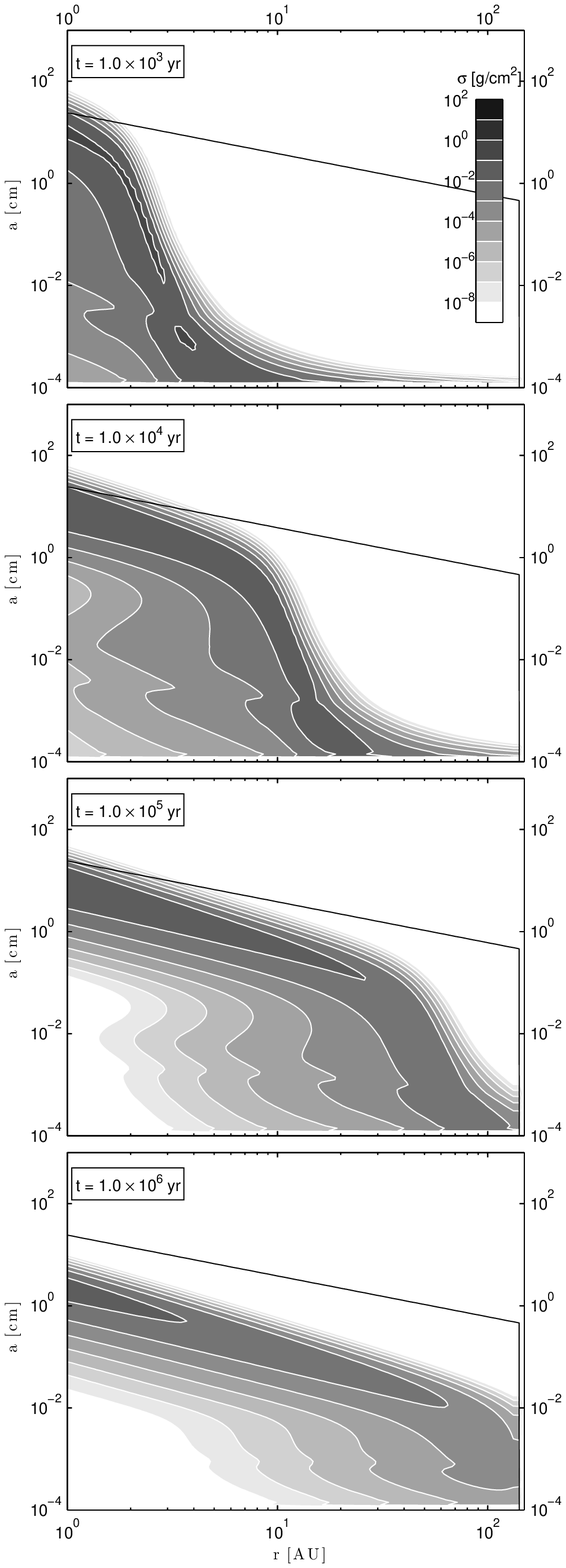}\includegraphics{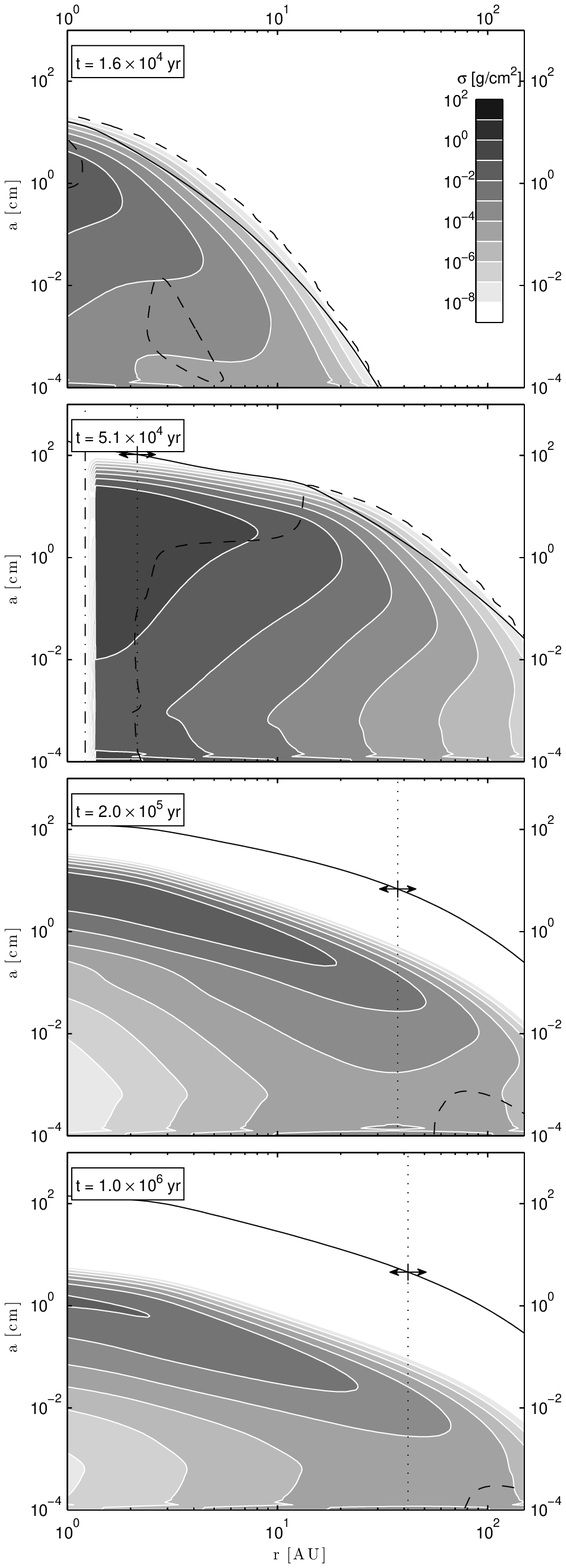}}
\caption{Snapshots of the vertically integrated dust density distributions (defined in Eq.~\ref{eq:def_sigma}) of a steady state disk as in \citetalias{Brauer:2008p215} (left column) and of an evolving disk (fiducial model, right column). No coagulation is calculated within the evaporation radius (denoted by the dash-dotted line), fragmentation is not taken into account in both simulations. The solid line shows the particle size corresponding to a Stokes number of unity.
Since $a_\mathrm{St=1} \propto \Siggas$ (see Eq.~\ref{eq:ST_epstein}) this curve
in fact has the same shape as $\Siggas(r)$, so it reflects, as a ``bonus'', what the gas disk looks like. The radius dividing the evolving disk into accreting and expanding regions is marked by the dotted line and the arrows. Particles which are located below the dashed line have a positive flux in the radial direction due to coupling to the expanding gas disk and turbulent mixing (particles within the closed contour in the upper right plot have an inward pointing flux).}
\label{fig:snapshots_fid}
\end{figure*}

\begin{table}
    \centering
    \caption[]{Parameters of the fiducial model.}
    \label{tab:model_parameters}
    $
    \begin{array}{p{0.5\linewidth}llp{0.1\linewidth}}
    \hline
    \noalign{\smallskip}
    parameter                       & \mathrm{symbol}   & \mathrm{value}    & unit\\
    \noalign{\smallskip}
    \hline
    \noalign{\smallskip}
    turbulence parameter            & \alpha            & 10^{-3}   & -\\
    irradiation angle               & \varphi           & 0.05      & -\\
    cloud mass                      & M_\text{cloud}    & 0.5       &$M_\odot$\\
    effect. speed of sound in core  & \cs               &3\e{4}     & cm~s$^{-1}$\\
    rotation rate of cloud core     & \Omega_\text{s}   &7\e{-14}   & s$^{-1}$\\
    solid density of dust grains        & \rhos         & 1.6       & g/cm$^3$\\
    stellar radius                  & R_\star           & 2.5       & $R_\odot$\\
    stellar temperature             & T_\star           & 4000      & K\\
    \noalign{\smallskip}
    \hline
    \end{array}
     $
\end{table}
We will now focus on the dust evolution of the disk. This fiducial simulation includes only grain growth without fragmentation or
erosion. All other parameters as well as the evolution of the gas surface
density and mid-plane temperature are the same as discussed in the previous section. The evolution of this model is visualized in Figure~\ref{fig:snapshots_fid}.

The contour levels in Figure~\ref{fig:snapshots_fid} show the vertically
integrated dust surface density distribution per logarithmic bin of grain
radius, $\sigma(r,a)$, as defined in Eq.~\ref{eq:def_sigma}.
The left column of Figure~\ref{fig:snapshots_fid} shows the results of dust
evolution for a steady state (i.e. not viscously evolving) gas disk as described
in \citetalias{Brauer:2008p215}.

The right column shows the evolution of the dust density distribution of the fiducial model, in which the gas disk is gradually built up through infall from the parent molecular cloud core, and the gas disk viscously spreads and accretes matter onto the star. The solid line marks the grain size corresponding to \St=1 at the given radius. This grain size will be called $a_{\St=1}$ hereafter.  In the Epstein regime, $a_{\St=1}$ is proportional to the gas surface density of the disk, which can be seen from Eq.~\ref{eq:ST_epstein}.

There are several differences to the simulations of grain growth in steady-state disks, presented in \citetalias{Brauer:2008p215}: viscously evolving disks accrete onto the star by transporting mass inwards and angular momentum outwards. This leads to the fact that some gas has to be moving to larger radii because it is 'absorbing' the angular momentum of the accreting gas. This can be seen in numerical simulations, but also in the self similar solutions of \citet{LyndenBell:1974p1945}. They show that there is a radius $R_\pm$ which divides the disk between inward and outward moving gas. This radius itself is constantly moving outwards, depending on the radial profile of the viscosity.
%,
%\begin{equation}
%R_\pm = R_1 \left( \frac{t/t_\text{s}+1}{2(2-\gamma)}\right)^{1/(2-\gamma)},
%\end{equation}
%where
%\begin{equation}
%t_\text{s} = \frac{1}{3(2-\gamma)^2} \frac{R_1^2}{\nu_1}
%\end{equation}
%is the viscous scaling time, $R_1$ a radial scale factor and
%\begin{equation}
%\nu(R) = \nu_1 \left(\frac{R}{R_1}\right)^\gamma
%\end{equation}
%is the viscosity  \citep[for details, see][]{Hartmann:1998p664}.

The radius $R_\pm$ in the fiducial model was found to move from around 20~AU at the end of the infall to 42~AU at 1~Myr and is denoted by the dotted line in Figure~\ref{fig:snapshots_fid}.
Important here is that small particles are well enough coupled to the gas to be transported along with the outward moving gas while larger particles decouple from the gas and drift inwards. Those parts of the dust distribution which lie below the dotted line in Figure~\ref{fig:snapshots_fid} have positive flux in the radial direction due to the gas-coupling.

One might expect that the dotted and dashed lines always coincide for small grains as they are well coupled to the gas, however, it can be seen that this is not the case. The reason for this is that turbulent mixing also contributes to the total flux of dust of each grain size. The smallest particles in between the dotted line and the dashed line in the lower two panels of Figure~\ref{fig:snapshots_fid} do have positive velocities, but due to a gradient in concentration of these dust particles, the flux is still negative.

During the early phases of its evolution, a disk which is built up from inside out quickly grows large particles at small radii which are lost to the star by radial drift. During further evolution, growth timescales become larger and larger (since the dust-to-gas ratio is constantly decreasing) while only the small particles are mixed out to large radii.

\begin{figure*}[htb]
\centering
  \resizebox{0.9\hsize}{!}{\includegraphics{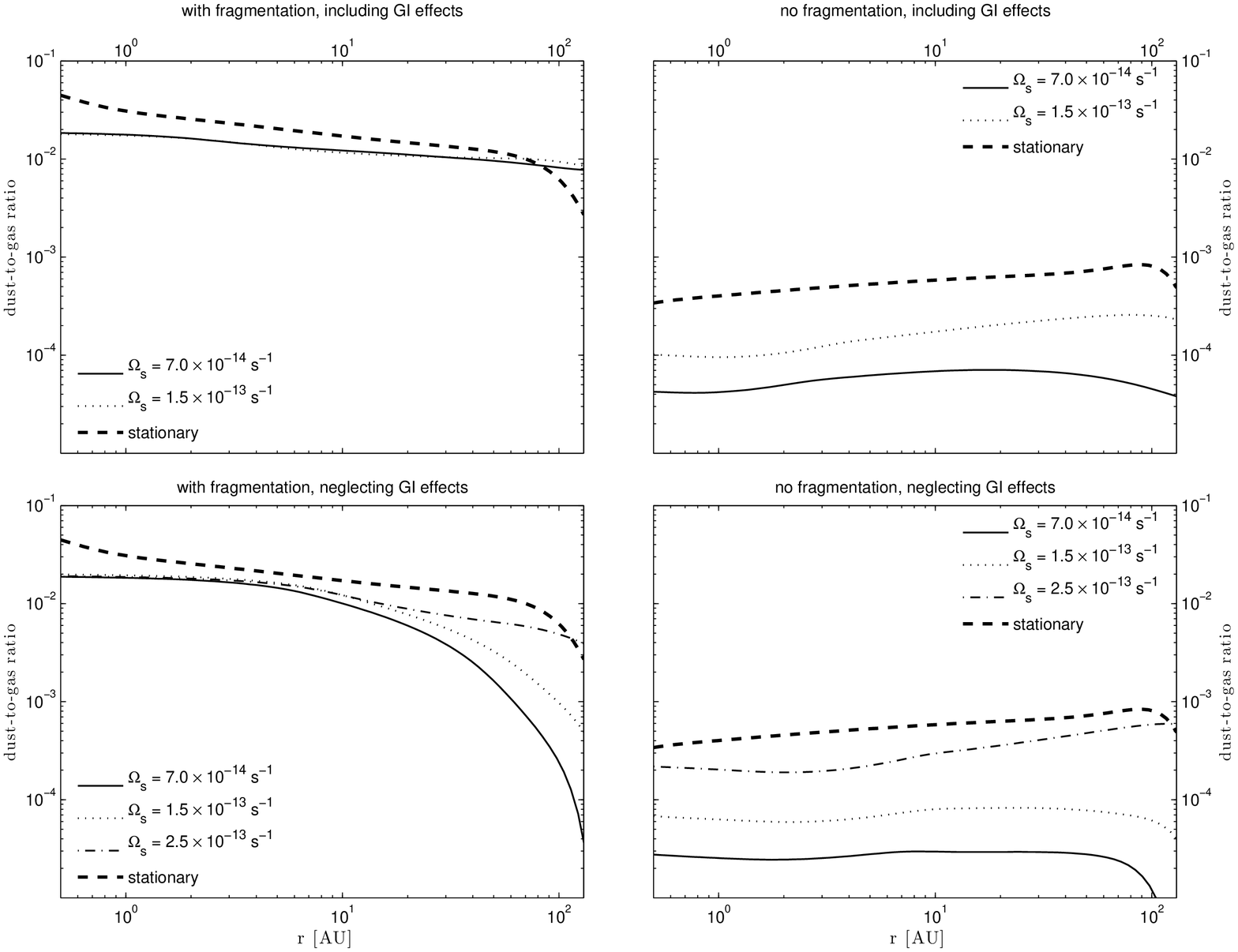}}
  \caption{\com{NEW FIGURE}Comparison of the radial dependence of the dust-to-gas ratio for the stationary simulations (thick lines) and the evolving disk simulations (thin lines). The four panels compare the results after 1~Myr of evolution with/without fragmentation (left/right column) and with/without effects of non-axisymmetric instabilities (top/bottom row). It can be seen that the dust-to-gas ratio of the evolving disk simulations is almost everywhere lower than the one of the stationary simulations. See Section~\ref{sec:influences_of_infall_model} for more details.}
  \label{fig:FGI}
\end{figure*}

The radial dependence of the dust-to-gas ratio after 1~Myr is shown in Figure~\ref{fig:FGI}.
These simulations show that the initial conditions of the stationary disk models (such as shown in \citetalias{Brauer:2008p215} and in the left column of Figure~\ref{fig:snapshots_fid}) are too optimistic since they assume a constant dust-to-gas ratio at the start of their simulation throughout the disk which is not possible if the disk is being built-up from inside out unless the centrifugal radius is very large (in which case the disk would probably fragment gravitationally) and grain fragmentation is included to prevent the grains from becoming large and start drifting strongly.
Removal of larger grains and outward transport of small grains lead to the fact that the dust-to-gas ratio is reduced by 0.5--1.5 orders of magnitude compared to a stationary model. This effect is also discussed in Section~\ref{sec:influences_of_infall_model}.

% ------------------------------------------------------------------------------
\subsection{Fiducial model with fragmentation}\label{sec:fiducial_model_frag}
The situation changes significantly, if we take grain fragmentation into account. As discussed in Section~\ref{sec:distr_of_fragments}, we consider two different kinds of outcomes for grain-grain collisions with relative velocities larger than the fragmentation velocity \uf: cratering (if the masses differ by less than one order of magnitude) and complete fragmentation (otherwise).

% \begin{figure}[htb]
% \centering
%   \resizebox{\hsize}{!}{\includegraphics{plots/fid_d2g}}
%   \caption{Radial dependence of the dust-to-gas ratio in the fiducial model without fragmentation (thin lines) and dust-to-gas ratio of a stationary disk model with comparable disk mass (thick dashed line). The times correspond to the snapshots shown in Figure~\ref{fig:snapshots_fid}.}
%   \label{fig:fid_d2g}
% \end{figure}

\begin{figure}[h!]
\centering
\makeatletter
\if@referee
	\resizebox{0.5\hsize}{!}{\includegraphics{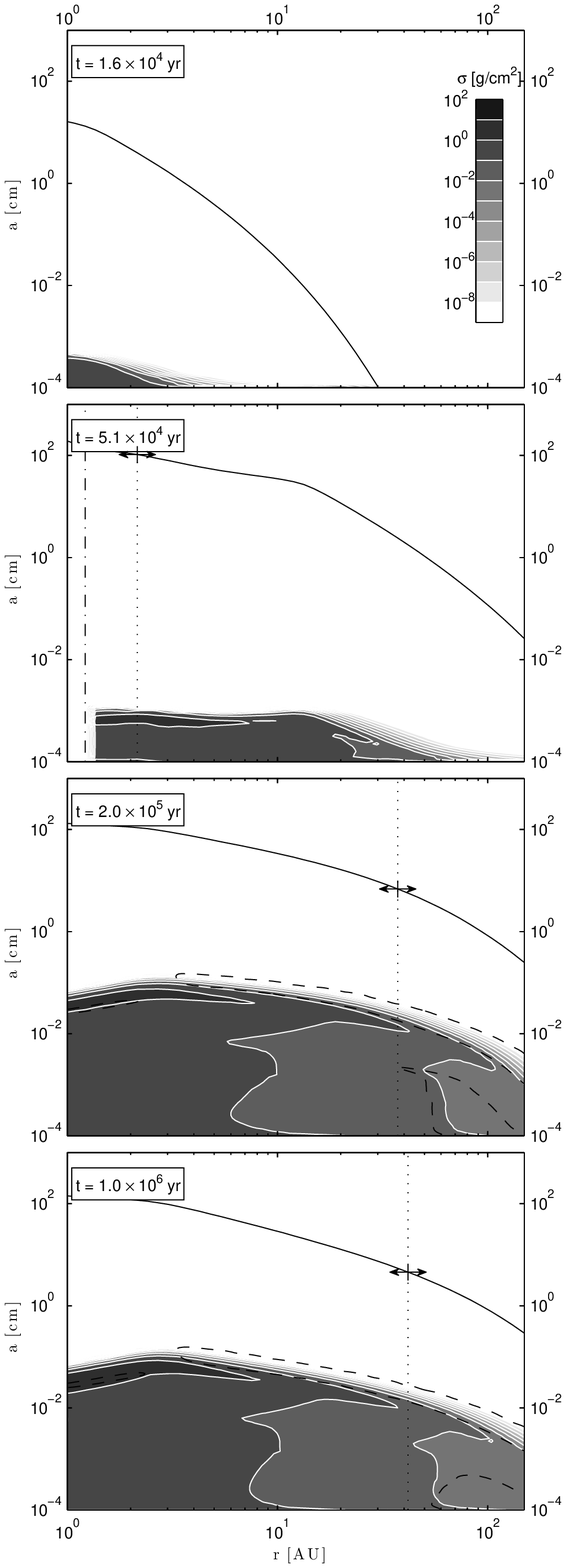}}
\else
	\resizebox{0.85\hsize}{!}{\includegraphics{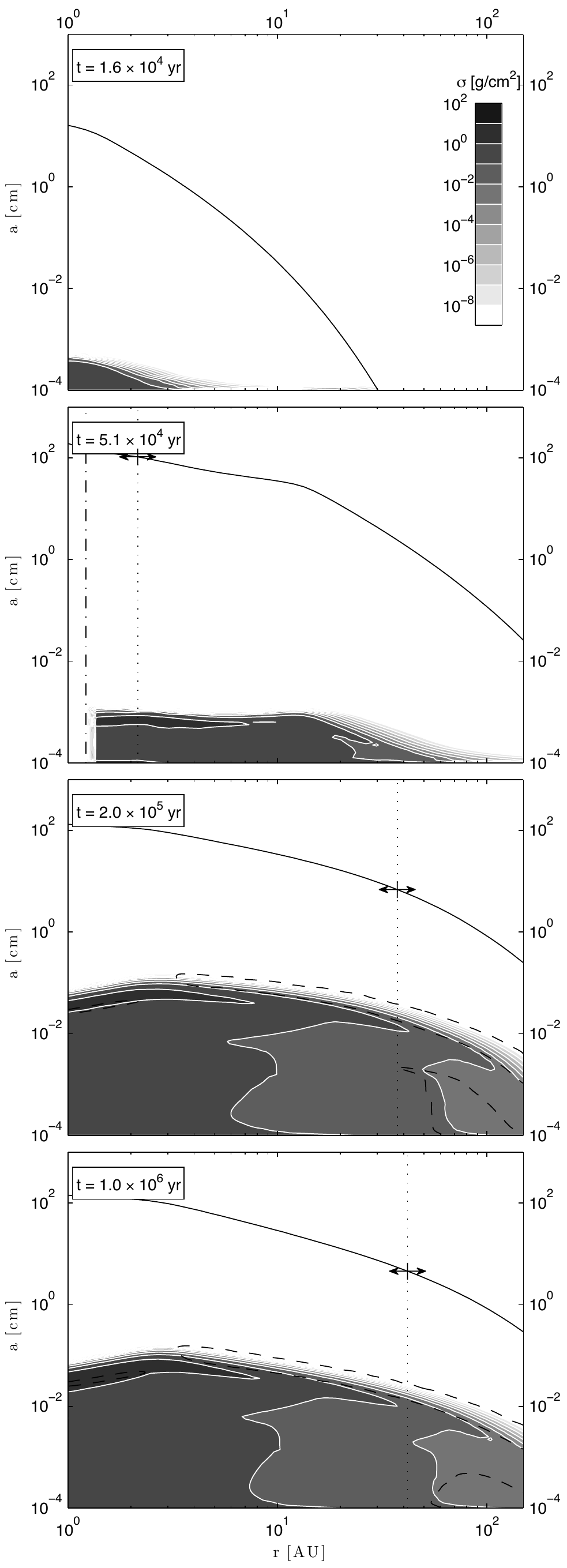}}
\fi
\makeatother
  \caption{Evolution of the dust density distribution of the fiducial model as in Figure~\ref{fig:snapshots_fid} but with fragmentation included. The dashed contour line (in the lower two panels) around the upper end of the size distribution and around small particles at $>60$~AU marks those parts of the distribution which have a positive (outward pointing) fluxes.}
  \label{fig:snapshots_frag}
\end{figure}

For particles larger than about $\St\approx 10^{-3}$, relative velocities are dominated by turbulent motion (and to a lesser extend by vertical settling). Since the relative velocities increase with Stokes number (and therefore with grain size), we can calculate the approximate position of the fragmentation barrier by equating the assumed fragmentation velocity \uf with the approximate relative velocities of the particles,
\begin{equation}
\St_\text{max} \simeq \frac{\uf^2}{2\alpha\,\cs^2}
\label{eq:st_max}
\end{equation}
Particles larger than this size are subject to high-velocity collisions which will erode or completely fragment those particles. This is only a rough estimate as the relative velocities also depend on the size of the smaller particle and radial drift also influence the grain distribution. However Eq.~\ref{eq:st_max} reproduces well the upper end of the size distribution in most of our simulations and therefore helps understanding the influence of various parameters on the outcome of these simulations.

The evolution of the grain size distribution including fragmentation is depicted in Figure~\ref{fig:snapshots_frag}. The initial condition is quickly forgotten since particles grow on very short timescales to sizes at which they start to fragment. The resulting fragments contribute again to the growth process until a semi-equilibrium of growth and fragmentation is reached.

\begin{figure}[tbh]
  \resizebox{\hsize}{!}{\includegraphics{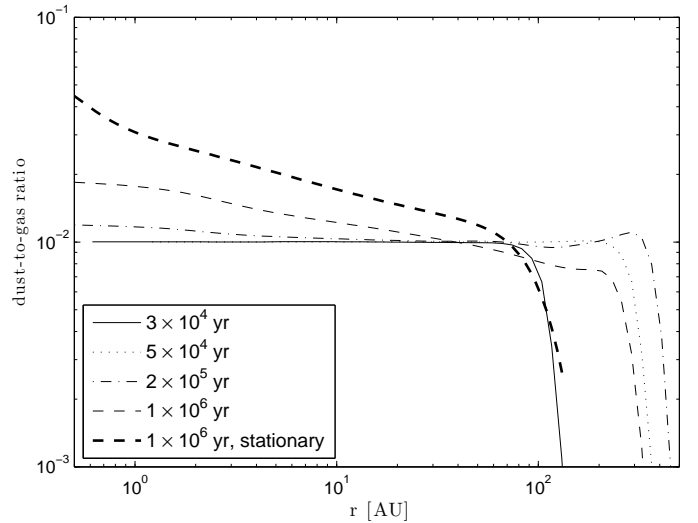}}
  \caption{Evolution of the radial dependence of the dust-to-gas ratio in the fiducial model including fragmentation with the times corresponding to the snapshots shown in Figure~\ref{fig:snapshots_frag}. The initial dust-to-gas ratio is taken to be~0.01. The thick dashed curve represents the result at 1~Myr of the static disk model for comparison.}
  \label{fig:fid_d2g_frag}
\end{figure}

It can be seen that particles stay much smaller than in the model without fragmentation. This means that 
they are less affected by radial drift on the one hand and better transported along with the expanding gas disk on the other hand. Consequently, considerable amounts of dust can reach radii of the order of 100~AU, as seen in Figure~\ref{fig:fid_d2g_frag}.

The approximate maximum Stokes number, defined in Eq.~\ref{eq:st_max}, is inversely proportional to the temperature (since relative velocities are proportional to \cs), which means that in regions with lower temperature, particles can reach larger Stokes numbers. By equating drift and drag velocities of the particles (cf. Eq.~\ref{eq:u_r_dust}), it can be shown that the radial velocities of particles with Stokes numbers larger than about $\alpha/2$, are being dominated by radial drift.

Due to the high temperatures below $\sim$5~AU (caused by viscous heating), $\St_\text{max}$ stays below this value which prevents any significant radial drift within this radius, particles inside 5~AU are therefore only transported along with the accreting gas. Particles at larger radii and lower temperatures can drift (although only slightly), which means that there is a continuous transport of dust from the outer regions into the inner regions. Once these particles arrive in the hot region, they get ``stuck'' because their Stokes number drops below $\alpha/2$. The gas within about 5~AU is therefore enriched in dust, as seen in Figure~\ref{fig:fid_d2g_frag}. The dust-to-gas ratio at 1~AU after 1~Myr is increased by 25\% over the value of in-falling matter, which is taken to be 0.01.

Figure~\ref{fig:fid_d2g_frag} also shows a relatively sharp decrease in the dust to gas ratio at a few hundred AU. At these radii, the gas densities become so small that even the smallest dust particles decouple from the gas and start to drift inwards.

The thick line in Figure~\ref{fig:fid_d2g_frag} shows as comparison the dust-to-gas ratio of the stationary disk model (cf. left column of Figure~\ref{fig:snapshots_fid}) after 1~Myr, which starts with a radially constant initial dust-to-gas ratio of 0.01.

\begin{figure}[thb]
  \resizebox{\hsize}{!}{\includegraphics{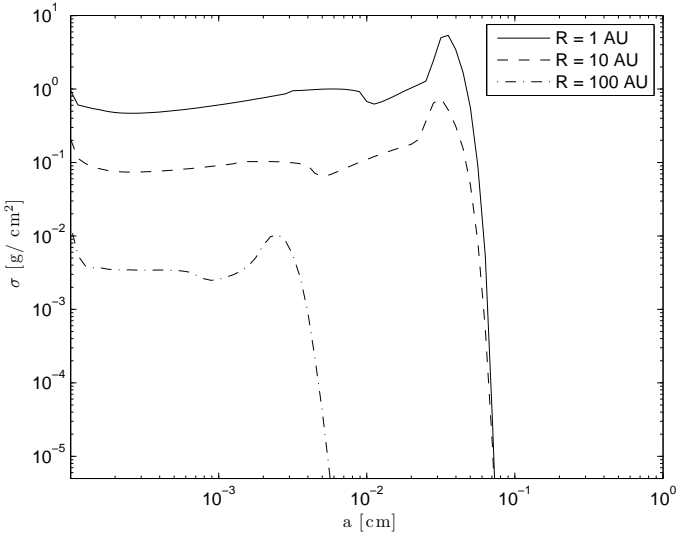}}
  \caption{Vertically integrated (cf. Eq.~\ref{eq:def_sigma}) grain surface density distributions as function of grain radius at a distance of 1~AU (solid), 10~AU (dashed) and 100~AU (dot-dashed) from the star. These curves represent slices through the bottom panel of Figure~\ref{fig:snapshots_frag}.}
  \label{fig:distri_slices}
\end{figure}

Figure~\ref{fig:distri_slices} shows the semi-equilibrium grain surface density distribution at 1, 10 and 100~AU in the fiducial model with fragmentation at 1~Myr.
The exact shape of these distributions depends very much on the prescription of fragmentation and cratering. In general the overall shape of these semi-equilibrium distributions is always the same: a power-law or nearly constant distribution (in $\sigma$) for small grains and a peak at some grain size $a_\mathrm{max}$, beyond which the distribution drops dramatically. The peak near the upper end of the distribution is caused by cratering. This can be understood by looking at the collision velocities: the relative velocity of two particles increases with the grain size but it is lower for equal-sized collisions than for collisions with particles of very different sizes (see Figure~\ref{fig:rel_vel}). The largest particles in the distribution have relative velocities with similar sized particles which lie just below the fragmentation velocity (otherwise they would fragment). This means that the relative velocities with much smaller particles (which are too small to fragment the bigger particles but can still damage them via cratering) are above this critical velocity. This inhibits the further growth of the big particles beyond $a_\mathrm{max}$, causing a ``traffic jam'' close to the fragmentation barrier. The peak in the distribution represents this traffic jam.

%-----------------------------------------------------------------------
\subsection{Influences of the infall model}\label{sec:influences_of_infall_model}
In the fiducial model without fragmentation, continuous resupply of material by infall is the cause why the disk has much more small grains than compared to the stationary disk model (cf. Figure~\ref{fig:snapshots_fid}), which relatively quickly consumes all available micrometer sized dust. The effect has already been found in \citet{Dominik:2008p4626}: if all grains start to grow at the same time, then the bulk of the mass grows in a relatively thin peak to larger sizes (see Figure~6 in \citetalias{Brauer:2008p215}). However if the bulk of the mass already resides in particles of larger size, then additional supply of small grains by infall is not swept up effectively because of the following: firstly, the number density of large particles is small (they may be dominating the mass, but not necessarily the number density distribution) and secondly, they only reside in a thin mid-plane layer while the scale height of small particles equals the gas scale height.

We studied how much the disk evolution depends on changes in the infall model.

For a given cloud mass, the so called centrifugal radius $r_\text{centr}$, which was defined in Eq.~\ref{eq:r_centri}, depends on the temperature and the angular velocity of the cloud. Both can be varied resulting in a large range of possible centrifugal radii reaching from a few to several hundred AU. Since the centrifugal radius is the relevant parameter, we varied only the rotation rate of the cloud core. We performed simulations with three different rotation rates which correspond to centrifugal radii of about 8 (fiducial model), 33~AU, and 100~AU. For each centrifugal radius, we performed two simulations: one which includes effects of gravitational instabilities (GI) -- i.e. increased $\alpha$ during infall and according to Eq.~\ref{eq:alpha_of_Q} -- and one which neglects them.

However for a centrifugal radius of 100~AU, too much matter is loaded onto the cold outer parts of the disk and consequently, the disk would fragment through gravitational instability. We cannot treat this in our simulations, hence, for the case of 100~AU, we show only results which neglect all GI effects.

The resulting dust-to-gas ratios are being shown in Figure~\ref{fig:FGI}.

Two general aspects change in the case of higher rotation rates: firstly, more of the initial cloud mass has to be accreted onto the star by going through the outer parts of the disk. Consequently, the disk is more extended and more massive than compared to the case of low rotation rate.

Secondly, as aforementioned the high surface densities in the colder regions at larger radii cause the disk to become less gravitationally stable.

If grain fragmentation is not taken into account in the simulations, both effects cause more dust mass to be transported to larger radii. Growth and drift timescales are increasing with radius and the dust disk with centrifugal radius of 33~AU (100~AU) can stay 5 (35)~times more massive than in the low angular momentum case after 1~Myr if GI effects are neglected.

If GI effects are included, matter is even more effectively transported outward, the dust-to-disk mass ratio for 8 and 33~AU is increased from 5 to 8.

However if fragmentation is included, it does not matter so significantly, where the dust mass is deposited onto the disk since grains stay so small during the build-up phase of the disk (due to grain fragmentation by turbulent velocities) that they are well coupled to the gas. Outwards of  $\sim 10$~AU (without GI effects) or of a few hundred AU (if GI effects are included), the gas densities become so small that even the smallest grains start do decouple from the gas. They are therefore not as effectively transported outwards. In these regions, the amount of dust depends on the final centrifugal radius while at smaller radii, turbulent mixing quickly evens out all differences in the dust-to-gas ratio (see left column of Figure~\ref{fig:FGI}).

It can be seen, that in all simulations, the dust-to-gas ratio is lower than in the stationary disk model. The trend in the upper right panel in Figure~\ref{fig:FGI} suggests that for a centrifugal radius of 100~AU and the enhanced radial transport by GI effects might have a higher dust-to-gas ratio than the stationary disk model. However in this case, the disk would become extremely unstable and would therefore fragment.

The reason for this is the following: to be able to compare both simulations, the total mass of the disk-star system is the same as in the stationary disk models. How the total mass is distributed onto disk and star depends on the prescription of infall. If a centrifugal radius of 100~AU is used, the disk becomes so massive that it significantly exceeds the stability criterion $M_\mathrm{disk}/M_\star \lesssim 0.1$.

\begin{figure}[h]
  \centering
\makeatletter
\if@referee
	\resizebox{0.7\hsize}{!}{\includegraphics{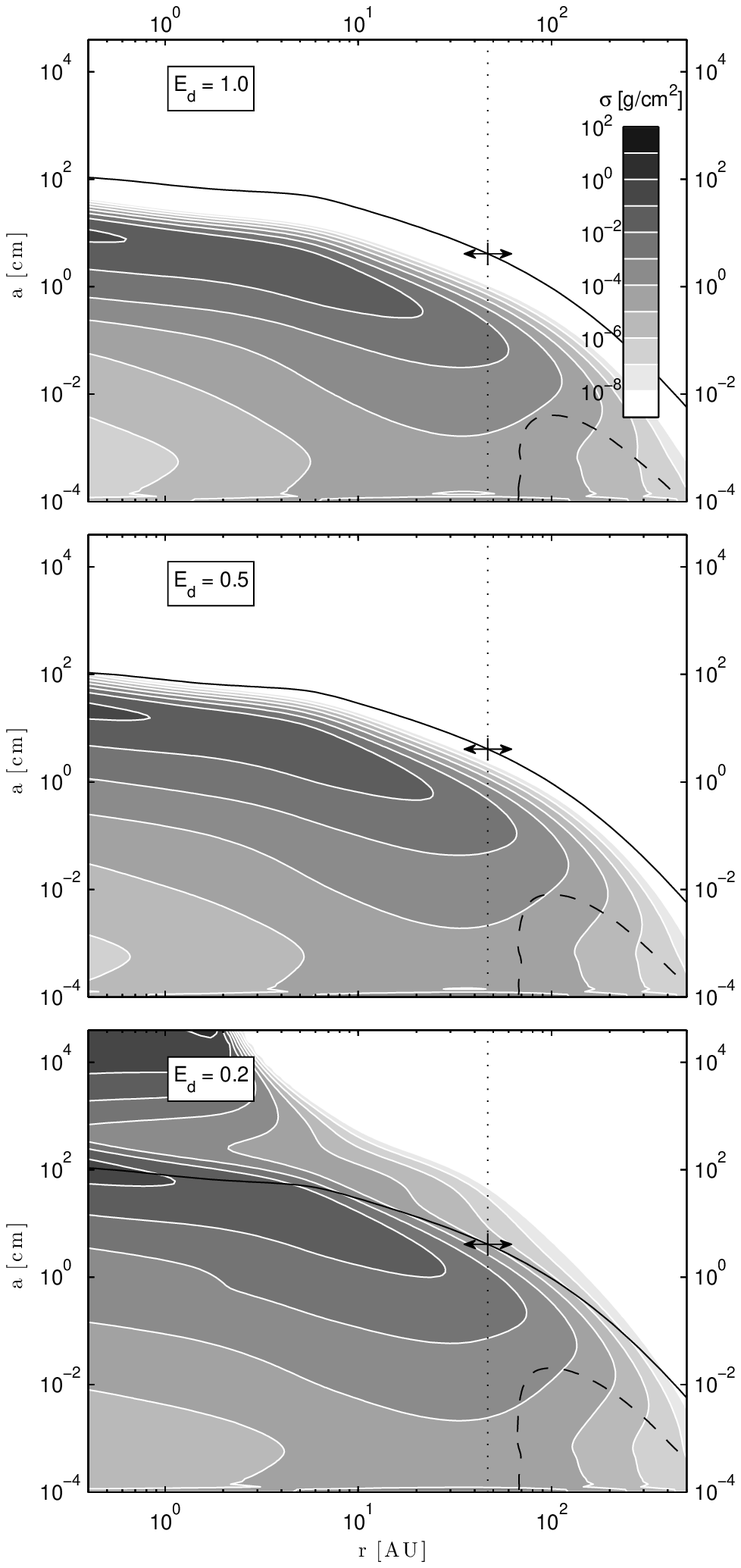}}
\else
	\resizebox{\hsize}{!}{\includegraphics{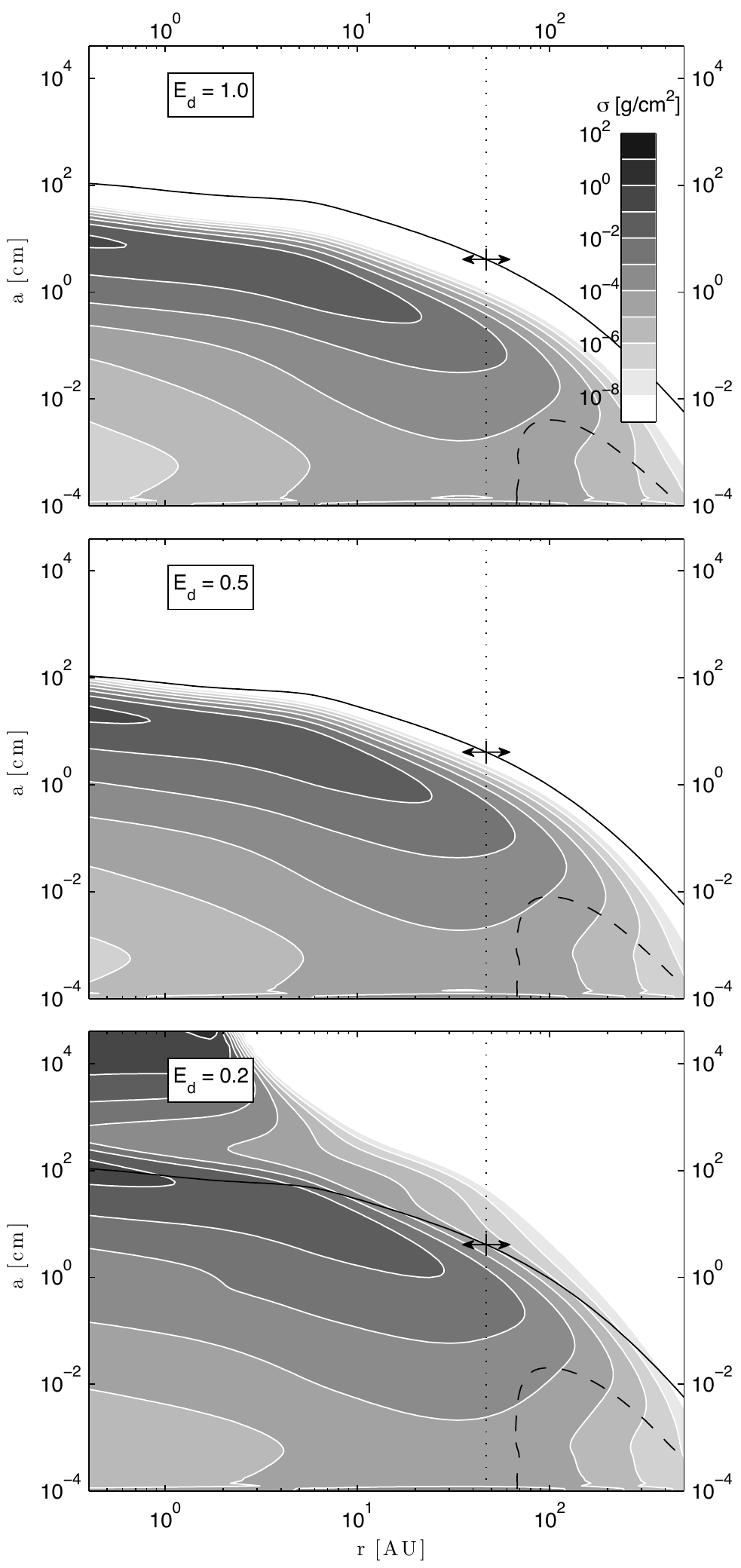}}
\fi
\makeatother
  \caption{Evolution of the dust surface density distribution of the fiducial model at 200\,000 years for different drift efficiencies $E_\text{d}$, without fragmentation. The solid line denotes the grain size $a_{\St=1}$ of particles with a Stokes number of unity. Gas outside of the radius denoted by the dotted line as well as particles below the dashed line have positive radial velocities. See section~\ref{sec:fiducial_model_nofrag} for more discussion.}
  \label{fig:snapshots_drift}
\end{figure}

%-----------------------------------------------------------------------
\subsection{The radial drift barrier revisited}\label{sec:drift_barrier}
According to the current understanding of planet formation, several mechanisms seem to prevent the formation of large bodies via coagulation quite rigorously. The most famous ones -- radial drift and fragmentation -- have already been introduced above.
Radial drift has first been discussed by \citet{Weidenschilling:1977p865}, while the importance of the fragmentation barrier (which may prevent grain growth at even smaller sizes) was discussed in \citetalias{Brauer:2008p215}.
In the following, we want to test some ideas about how to weaken or to overcome these barriers apart from those already studied in \citetalias{Brauer:2008p215}.

\citetalias{Brauer:2008p215} has quantified the radial drift barrier by equating the timescales of growth and radial drift which leads to the condition
\begin{equation}
\frac{\tau_\text{g}}{\tau_\text{d}} = \frac{1}{\epsilon_0} \left( \frac{\Hp}{r} \right)^2 \leq \frac{1}{\gamma},
\label{eq:drift_barrier}
\end{equation}
where $\epsilon_0$ is the dust-to-gas ratio and $\tau_\text{d}$ and $\tau_\text{g}$ are the drift and growth timescales respectively. The  parameter $\gamma$ describes how much more efficient growth around \St=1 must be, so that the particles are not removed by radial drift. To overcome the drift barrier, obviously either particle growth must be accelerated, or the drift efficiency has to be decreased. \citetalias{Brauer:2008p215} have numerically measured the parameter $\gamma$ to be around 12. In other words, this means that the growth timescales have to be decreased (e.g. by an increased dust-to-gas ratio) until the condition in Eq.~\ref{eq:drift_barrier} is fulfilled.

However, there are other ways of breaking through the drift barrier. Firstly, the drift timescale for \St=1 particles also depends on the temperature (via the pressure gradient). A simple approximation from Eq.~\ref{eq:drift_barrier} with a 0.5~$M_\odot$ star and a dust-to-gas ratio of 0.01 gives
\begin{equation}
T < 103\text{ K } \left(\frac{r}{\text{AU}}\right)^{-1},
\end{equation}
which means that particles should be able to break through the drift barrier at 1~AU if the temperature is below $\sim$100~K (or 200~K for a solar mass star).
\citet{Dullemond:2002p399} have constructed vertical structure models of passively irradiated circumstellar disks using full frequency- and angle-dependent radiative transfer. They show that the mid-plane temperature of such a T Tauri like system at 1~AU can be as low as 60~K. Reducing the temperature by some factor reduces the drift time scale by the a factor of similar size which we will call the radial drift efficiency $E_\text{d}$ (cf. Eq.~\ref{eq:u_eta}).

Zonal flows as presented in \citet{Johansen:2006p7466} could be an alternative way of decreasing the efficiency of radial drift averaged over typical time scales of particle growth. \citet{Johansen:2006p7466} found a reduction of the radial drift velocity of up to 40\% for meter-sized particles.

Meridional flows \citep[e.g.,][]{Urpin:1984p1473,Kley:1992p7134} might also seem interesting in this context, however they do not directly influence the radial drift efficiency but rather reverse the gas-drag effect. This might be important for small particles (which, however are not strongly settling to the mid-plane) but for \St=1 particles, $\alpha$ would have to be extremely high to have significant influence: even $\alpha=0.1$ would result in a reduction of the particles radial velocity by approximately only a few percent.

Another possibility to weaken the drift barrier is changing its radial dependence. The reason for this is the following: particle radial drift is only a barrier if it prevents particles to cross the size $a_{\St=1}$. Since particles grow while drifting, the particle size corresponding to \St=1 needs to increase as well, to be a barrier. Otherwise drifting particles would grow (at least partly) through the barrier while they are drifting. If $a_{\St=1}$ is decreasing in the direction toward the star, then a particle that drifts inwards would have an increasing Stokes number even if the particle does not grow at all.

\begin{figure}[thb]
  \vspace{0.35cm}
  \centering
  \resizebox{\hsize}{!}{\includegraphics{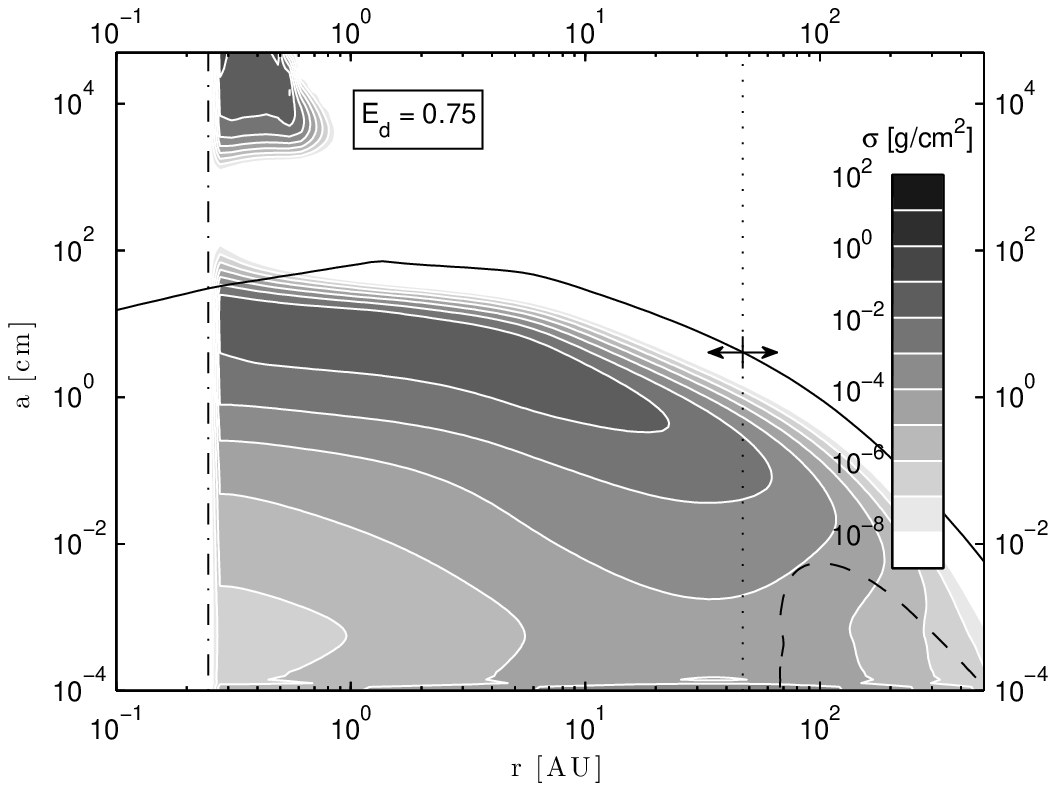}}
  \caption{Dust grain surface density distribution as in Figure~\ref{fig:snapshots_drift} at 200\,000 years but including the Stokes drag regime. The drift efficiency is set to $E_\text{d} = 0.75$ and fragmentation is not taken into account. It can be seen that $a_\mathrm{\St=1}$ (solid line) increases with radius until about 1~AU, which facilitates the break through the drift barrier.}
  \label{fig:snapshots_drift_ST}
\end{figure}

In the Epstein regime, the size corresponding to \St=1 is proportional to the gas surface density
\begin{equation}
a_\text{\St=1} = \frac{2 \Siggas}{\pi \rhos},
\end{equation}
meaning that a relatively flat profile of surface density (or even a profile with positive slope) is needed to allow particles to grow through the barrier. However, our simulations of the viscous gas disk evolution does not yield surface density profiles with positive slopes outside the dust evaporation radius.

To quantify the arguments above, we have performed simulations with varying drift efficiency $E_\text{d}$ to test how much the radial drift has to be reduced to allow break through. We have additionally included the first Stokes drag regime to see how the radial drift of particles is influenced by it.

Figure~\ref{fig:snapshots_drift} shows the grain surface density distribution after 200\,000 years of evolution for three different drag efficiencies. The most obvious changes can be seen in the region where the $a_{\St=1}$ line (solid line) is relatively flat: the grain distribution is shifted towards larger Stokes numbers. As explained above, particles can grow while drifting, which can be seen in the case of $E_\text{d}=0.5$. The Stokes number and size of the largest particles is significantly increasing towards smaller radii. However the radial drift efficiency has to be reduced by 80\% to produce particles which are large enough to escape the drift regime and are therefore not lost to the star.

\begin{figure*}[thb]
\centering
\vspace{0.45cm}
\resizebox{\hsize}{!}{\includegraphics{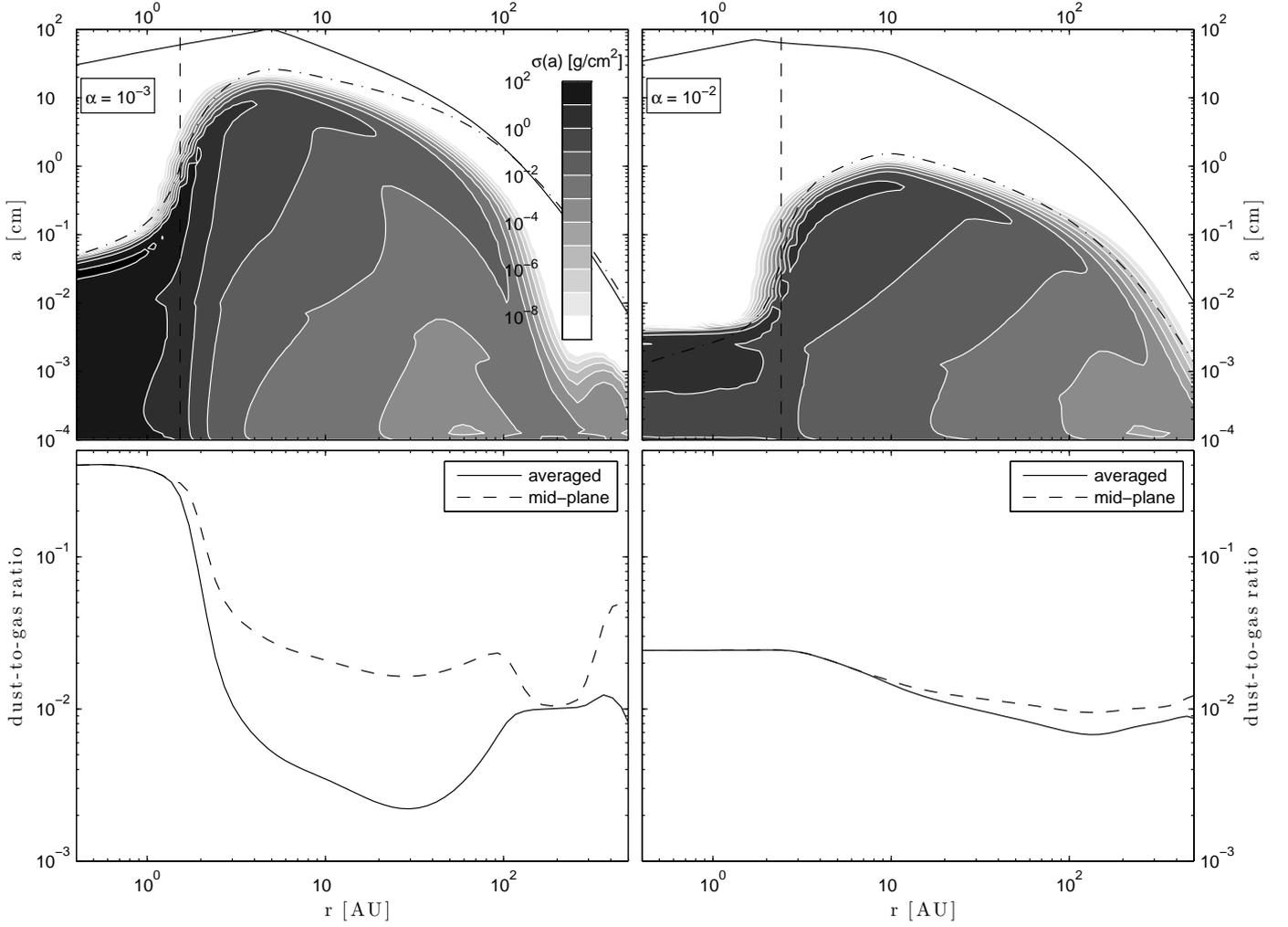}}
\caption{Dust surface density distributions (top row) and the according solid-to-gas ratio (bottom row) for the case of radius-dependent fragmentation velocity after 2\e{5} years of evolution. In the upper row, the vertical dashed line denotes the position of the snow line at the mid-plane, the solid line corresponds to $a_\mathrm{St=1}$ and the dot-dashed line shows the approximate location of the fragmentation barrier according to Eq.~\ref{eq:st_max}.
The snow line on the right plot lies further outside since viscous heating is stronger if $\alpha$ is larger.
In the bottom row, the solid line denotes the vertically integrated dust-to-gas ratio while the dashed line denotes the dust-to-gas ratio at the disk mid-plane.
The icy dust grains outside the snow line are assumed to fragment at a critical velocity of 10~m~s$^{-1}$ while particles inside the snow line fragment at 1~m~s$^{-1}$. The plots on the left and right hand side differ in the amount of turbulence in the disk ($\alpha = 10^{-3}$ and $10^{-2}$, respectively).}
\label{fig:snapshots_enhancement}
\end{figure*}

The situation changes, if the Stokes drag is taken into account: if gas surface densities are high enough for the dust particles to get into a different drag regime, a change in the radial dependency of $a_\text{\St=1}$ can be achieved. The Epstein drag regime for particles with Stokes number of unity is only valid if
\begin{equation}
\Siggas \lesssim 108\frac{\text{g}}{\text{cm}^2} \left(\frac{T}{\text{200 K}}\right)^{\frac{1}{4}} \left(\frac{R}{\text{AU}}\right)^{\frac{3}{4}}   \left(\frac{M_\star}{M_\odot}\right)^{-\frac{1}{4}}\left(\frac{\rhos}{1.6\frac{\text{g}}{\text{cm}^3}}\right)^{\frac{1}{2}},
\label{eq:sig_threshold}    
\end{equation}
otherwise, drag forces have to be calculated according to the Stokes drag law since the Knudsen number becomes smaller than 4/9 \citep[see][]{Weidenschilling:1977p865}. The Stokes number is then given by
\begin{equation}
\St = \frac{\sqrt{2\pi}}{9} \frac{\rhos\; \sigma_{\text{H}_2} \; a^2}{\mu\; \mp} \: \Hp^{-1},
\end{equation}
with $\sigma_{\text{H}_2}$ being the cross section of molecular hydrogen. Interestingly, $a_{\St=1}$ is independent of $\Siggas$ and proportional to the square root of the pressure scale height which decreases towards smaller radii. This means that -- as long as the surface density is high enough -- it does not depend on the radial profile of the surface density. In this regime the radial drift itself could move particles over the drift barrier since drifting inwards increases the Stokes number of a particle without increasing its size.

Results of simulations which include the Stokes drag are shown in Figure~\ref{fig:snapshots_drift_ST}. In this case, particles can already break through the drift barrier if $E_\text{d} \lesssim 0.75$. This value and the position of the breakthrough depends on where the drag law changes from Epstein to Stokes regime and therefore on the disk surface density. As noted above, larger surface densities generally shift the position of regime change towards larger radii making it easier for particles to break through the drift barrier.

It should be noted that the physical way to avoid the Stokes drag regime is to decrease the surface densities, however we chose the same initial conditions for both cases -- with and without Stokes drag -- and just neglected the Stokes drag in the latter computations to be able to compare the efficiency factors independent of other parameters such as disk mass or temperature.

% ------------------------------------------------------------------------------
\subsection{The fragmentation barrier revisited}\label{sec:frag_barrier}

In the previous section, we have shown that several mechanisms could allow particles to break through the radial drift barrier, however fragmentation puts even stronger constraints on the formation of planetesimals.

As shown by \citet{Ormel:2007p801}, the largest relative velocities are of the order of
\begin{equation}
\Delta u_\text{max} \simeq \sqrt{2 \, \alpha} \, \cs.
\end{equation}
If particles should be able to break through the fragmentation barrier, then they need to survive these large relative velocities, meaning that $\Delta u_\text{max}$ has to be smaller than the fragmentation velocity of the particles, or
\begin{equation}
\frac{\uf}{\cs} \gtrsim \sqrt{2 \, \alpha}.
\end{equation}
This condition is hard to fulfill with reasonable fragmentation velocities, unless $\alpha$ is very small. E.g., for $\alpha = 10^{-5}$ and a temperature of 200 to 250~K, the fragmentation velocity needs to be higher than 4~m~s$^{-1}$, which could already seen in the simulations by \cite{Brauer:2008p212}, who have simulated particle growth near the snow-line.

Even in the case of very low turbulence, relative azimuthal velocities of large ($\St\gtrsim 1$) and small grains ($\St\ll 1$) are of the order of 30~m~s$^{-1}$, which means that large particles are constantly being 'sand-blasted' by small grains. The only way of reducing these velocities significantly is decreasing the pressure gradient (see Equations~\ref{eq:u_eta}~and~\ref{eq:dv_az}).

Another possibility to overcome this problem would be if high-velocity impacts of smaller particles would cause net growth, as has been found experimentally by \citet{Wurm:2005p1855} and \citet{Teiser:2009p7785}.

Taken together, these facts make environments as the inner edge of dead zones ideal places for grain growth \citep[see][]{Brauer:2008p212,Kretke:2007p697}: shutting of MRI leads to low values of $\alpha$, which are needed to reduce turbulent relative velocities and the low pressure gradients prevent radial drift and high azimuthal relative velocities.

% ------------------------------------------------------------------------------
\subsection{Dust enhancement inside the snow line}\label{sec:dustenh_in_snowline}

As already noted in a previous paper \citep{Birnstiel:2009p7135}, significant loss of dust by radial drift can be prevented by assuring that particles stay small enough and are therefore not influenced by radial drift. For typical values of $\alpha$ ($10^{-3}-10^{-2}$), this means that the fragmentation velocity must be smaller than about 0.5--5~m~s$^{-1}$. If particles have higher tensile strength, they can grow to larger sizes which are again affected by radial drift.

Typical fragmentation velocities for silicate grains determined both theoretically and experimentally are of the order of a few~m~s$^{-1}$ \citep[for a review, see][]{Blum:2008p1920}. The composition of particles outside the snow-line is expected to change due to the presence of ices. This can influence material properties and increase the fragmentation velocity \citep[see][]{Schafer:2007p7468,Wada:2009p8776}.

We have performed simulations with a radially varying fragmentation speed. We assume the fragmentation speed inside the snow-line to be 1~m~s$^{-1}$, outside the snow-line to be 10~m~s$^{-1}$. It should be noted that we do not follow the abundance of water in the disk or the composition of grains, we only assume particles outside the snow line to have larger tensile strength due to the presence of ice.
To be able to compare both simulations, we used the same 1~Myr old 0.09~$M_\odot$ gas disk around a solar mass star as initial condition.
The gas surface density profile of this disk was derived by a separate run of the disk evolution code. We used this gas surface density profile and a radially constant dust-to-gas ratio as initial condition for the simulations presented in this subsection. Apart from the level of turbulence, the input for both simulations is identical, the results are therefore completely independent of uncertainties caused by the choice of the infall model.

Results of the simulations are shown in Figure~\ref{fig:snapshots_enhancement}.
A one order of magnitude higher fragmentation velocity causes the maximum grain size to be about two orders of magnitude larger, which follows from Eq.~\ref{eq:st_max} since $\St_\text{max} \propto a_\text{max}$ in the Epstein regime (all particles in these simulations are small enough to be in the Epstein regime). This effect can be seen in Figure~\ref{fig:snapshots_enhancement}.
Particles outside the snow-line are therefore more strongly drifting inwards (because they reach larger Stokes numbers) where they are being pulverized as soon as they enter the region within the snow-line.

Strong drift outside the snow line and weaker radial drift inside the snow line cause the dust-to-gas ratio within the snow line to increase significantly (see bottom row of Figure~\ref{fig:snapshots_enhancement}): in the case of $\alpha = 10^{-3}$, the dust-to-gas ratio reaches values between 0.39 and 0.10 in the region from 0.2 to 1.9~AU.

Simulations for a less massive star (0.5~$M_\odot$) show the same behavior, however the increase in dust-to-gas ratio is not as high as for a solar mass star (dust-to-gas ratio of 0.27--0.20 from 0.2--4~AU).

\begin{figure}[thb]
  \centering
  \vspace{0.45cm}
  \resizebox{\hsize}{!}{\includegraphics{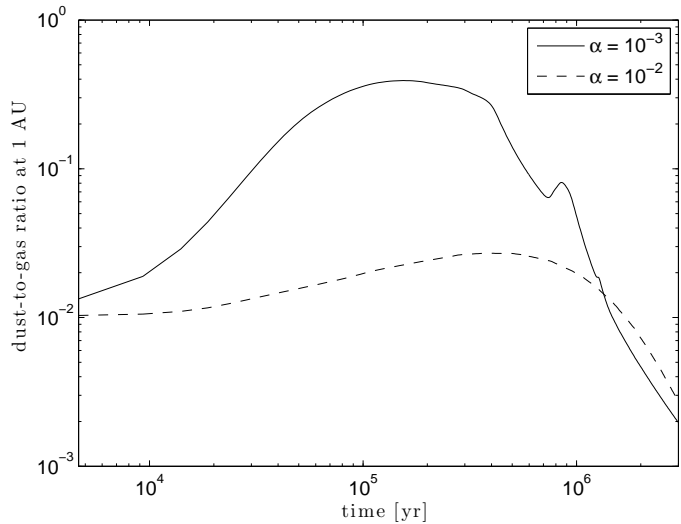}}
  \caption{Dust-to-gas ratio at a distance of 1~AU from the central star as a function of time for the case of low ($\alpha=10^{-3}$, solid line) and high ($\alpha=10^{-2}$, dashed line) turbulence.}
  \label{fig:d2g_at_1AU}
\end{figure}

This effect is not as significant in the case of stronger turbulence, where the maximum dust-to-gas ratio is around 0.027. The evolution of the dust-to-gas ratio at a distance of 1~AU from the star is plotted for both cases in Figure~\ref{fig:d2g_at_1AU} (the minor bump is an artifact of the initial condition).

The reason for this difference lies in the locations of the drift and fragmentation barriers. The approximate position of the fragmentation barrier is represented by the dot-dashed line in Figure~\ref{fig:snapshots_enhancement}. The radial drift barrier cannot be defined as sharply, however radial drift is strongest at a Stokes number of unity, which corresponds to the solid line in Figure~\ref{fig:snapshots_enhancement}. An increase of $\alpha$ by one order of magnitude lowers the fragmentation barrier by about one order of magnitude in grain size.

In the lower turbulence case, the fragmentation barrier lies close to $a_\mathrm{St=1}$. Most particles are therefore drifting inwards before they are large enough to experience the fragmentation barrier. Hence, the outer parts of the disk are significantly depleted in small grains.

In contrast to this case, fragmentation is the stronger barrier for grain growth throughout the disk in the high turbulence simulation (right column in Figure~\ref{fig:snapshots_enhancement}). It can be seen that particles of smaller sizes are constantly being replenished by fragmentation.

\begin{figure}[thb]
  \centering
  \vspace{0.45cm}
  \resizebox{\hsize}{!}{\includegraphics{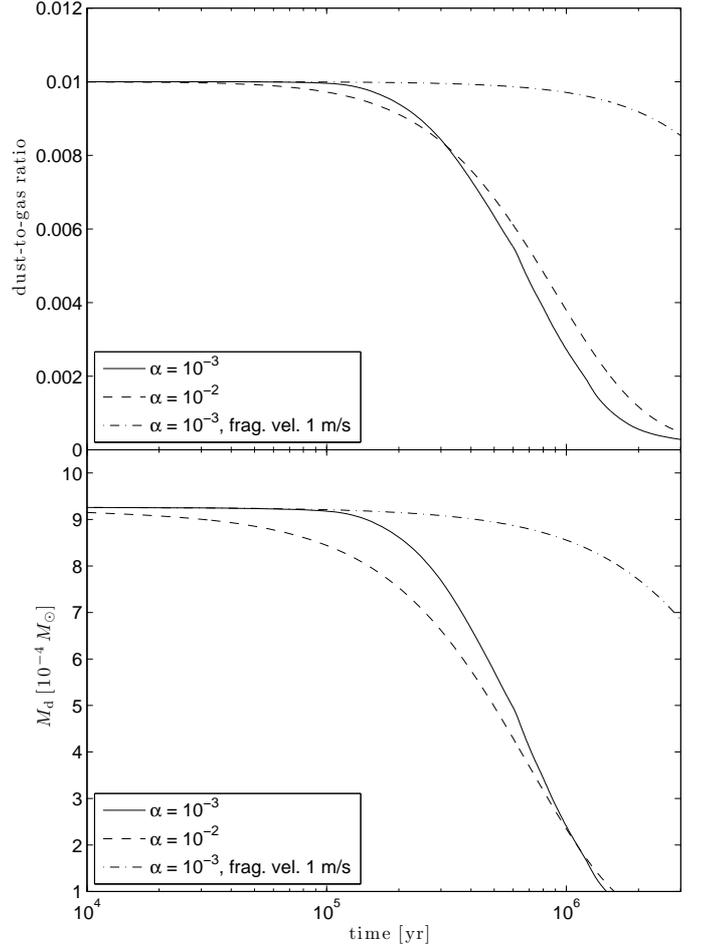}}
  \caption{Evolution of the global dust-to-gas ratio (top panel) and the dust disk mass (bottom panel) for the simulations shown in Figure~\ref{fig:snapshots_enhancement}. The solid and dashed lines correspond to the low and high turbulence case, respectively. The dash-dotted line shows the evolution of the disk if a low fragmentation velocity is assumed throughout the disk.}
  \label{fig:dust_mass_comparison}
\end{figure}

With these results in mind, the evolution of the disk mass (bottom panel of Figure~\ref{fig:dust_mass_comparison}) seems counter-intuitive: the mass of the high turbulence dust disk is decaying faster than in the low turbulence case. This can be understood by looking at the \emph{global} dust-to-gas ratio of the disks (top panel of Figure~\ref{fig:dust_mass_comparison}) which does not differ much in both cases. This means that the increased dust mass loss in the high turbulence disk is due to the underlying evolution of the gas disk. Particles in the high turbulence disk may have smaller Stokes numbers (causing drift to be less efficient), however the inward dragging by the accreting gas is stronger in this case.

To show how much the dust evolution depends on the fragmentation velocity, we included the case of a lower fragmentation velocity throughout the disk in Figure~\ref{fig:dust_mass_comparison}. It can be seen that the dust mass is retained at its initial value for much longer timescales.

% ==============================================================================
\section{Discussion and conclusions}\label{sec:discussion}

We constructed a new model for growth and fragmentation of dust in circumstellar disks.
We combined the (size and radial) evolution of dust of \citetalias{Brauer:2008p215} with a viscous gas evolution code which takes into account the spreading and accretion, irradiation and viscous heating of the gas disk. The dust model includes the growth/fragmentation, radial drift/drag and radial mixing of the dust. We re-implemented and substantially improved the numerical treatment of the Smoluchowski equation of \citetalias{Brauer:2008p215} to solve for the combined size and radial evolution of dust in a fully implicit, un-split scheme (see Appendix~\ref{app:alg}).
In addition to that, we also included more physics such as relative azimuthal velocities, radial dependence of fragmentation critical velocities and the Stokes drag regime.
The code has been tested extensively and was found to be very accurate and mass-conserving (see Appendix~\ref{app:tests}).

We compared our results of grain growth in evolving protoplanetary disks to those of steady state disk simulations, similar to \citetalias{Brauer:2008p215}. In spite of many differences in details, we confirm the most general result of \citetalias{Brauer:2008p215}: radial drift and particle fragmentation set strong barriers to particle growth. If fragmentation is included in the calculations, then it poses the strongest obstacle for the formation of planetesimals. Very low turbulence ($\alpha\lesssim 10^{-5}$) and fragmentation velocities of more than a few m~s$^{-1}$ are needed to be able to overcome the fragmentation barrier in the case of turbulent relative velocities.

This model includes also the initial build-up phase of the disk, which is still a very poorly understood phase of disk evolution. We use the Shu-Ulrich infall model which represents a strong simplification. However, the following novel findings of this work do not or only weakly depend on the build-up phase of the disk:
\begin{itemize}

\item Apart from an increased dust-to-gas ratio \citepalias[as in][]{Brauer:2008p215}, other mechanisms such as streaming instabilities or a decreased temperature may be able to weaken this barrier by decreasing the efficiency of radial drift. We found that in simulations without fragmentation the radial drift efficiency needs to be reduced by 80\% to produce particles which crossed the meter-size barrier and are large enough to resist radial drift.

\item If the gas surface density is above a certain threshold (in our simulations about 140~g~cm$^{-2}$ at 1~AU or 330~g~cm$^{-2}$ at 5~AU, see Eq.~\ref{eq:sig_threshold}) then the drag force which acts on the dust particles has to be calculated according to the Stokes drag law, instead of the Epstein drag law. The drift barrier in this drag regime is shifting to smaller sizes for smaller radii (independent on the radial profile of the surface density) which means that pure radial drift can already transport dust grains over the drift barrier or at least to larger Stokes numbers even without simultaneous grain growth. In this case, the drift efficiency needs to be reduced only by about 25\% to produce large bodies.

\item If relative azimuthal velocities are included, then grains with $\St>1$ are constantly 'sand-blasted' by small grains (if they are present) which (in our prescription of fragmentation) causes erosion and stops grain-growth even in the case of low turbulence. Only decreasing the radial pressure gradient significantly weakens both relative azimuthal and radial velocities. Low turbulence and a small radial pressure gradient together are needed to allow larger bodies to form. These conditions may be fulfilled at the inner edge of dead zones (\citealp{Brauer:2008p212}; \citealp{Kretke:2007p697}; see also Dzyurkevich et al., in press). Future work needs to investigate the disk evolution and grain growth of disks with dead zones.
Our prescription of fragmentation and erosion may also need rethinking since \citet{Wurm:2005p1855} and \citet{Teiser:2009p7785} find net-growth by high velocity impacts of small particles onto larger bodies.

\item Higher tensile strengths of particles outside the snow-line allows particles to grow to larger sizes, which are more strongly affected by radial drift. Particles therefore drift from outside the snow-line to smaller radii where they fragment and almost stop drifting (since their radial velocity is decreased by almost two orders of magnitude). This can cause an increase the dust-to-gas ratio inside the snow-line by more than 1.5 orders of magnitude.

\item The critical fragmentation velocity and its radial dependence (and to a lesser extent the level of turbulence) is a very important parameter determining if the dust disk is drift or fragmentation dominated. A drift dominated disk is significantly depleted in small grains and only a fragmentation dominated disk can retain a significant amount of dust for millions of years as is observed in T Tauri disks.
\end{itemize}

The following results depend on the build-up phase of the disk. However unless the collapse of the parent cloud is not inside-out or so fast to cause disk fragmentation, we expect only slight alteration of the results:
\begin{itemize}
\item Disk spreading causes small particles ($\lesssim 10$~$\mu$m) to be transported outward at radii of $\sim$60--190~AU even in 1~Myr old disks.

\item Small particles provided by infalling material are not effectively swept up by large grains if the bulk of the dust mass has already grown to larger sizes. 

\item In an inside-out build-up of circumstellar disks, grains growth is very fast (timescales of some 100~years) because densities are high and orbital timescales are small. Large grains are quickly lost due to drift towards the star if fragmentation is neglected. Fragmentation is firstly needed to keep grains small enough to be able to transport a significant amount of dust to large radii by disk spreading and secondly to retain it in the disk by preventing strong radial inward drift.

\end{itemize}

\begin{acknowledgements}
We like to thank Thomas Henning, Hubert Klahr, Chris Ormel, and Andras Zsom for insightful discussions. We also thank the referee, Hidekazu Tanaka for his fast and insightful review which helped to improve this paper.
\end{acknowledgements}

% ==============================================================================

\bibliographystyle{aa}
\bibliography{/Users/til/Documents/Papers/bibliography}

% ==============================================================================
\makeatletter
\if@referee
\processdelayedfloats
\fi
\makeatother
% ===============================================
\appendix 
% ------------------------------------------------------------------------------
\section{Algorithm}\label{app:alg}
% ------------------------------------------------------------------------------
In the next two sections, we will first discuss how the equations of radial evolution of gas (Eq. \ref{eq:ssd}) and dust (Eq. \ref{eq:dustequation}) as well as the coagulation/fragmentation of dust (Eq. \ref{eq:smolu}) are solved separately. In Section \ref{app:alg_both} we will then describe how this model treats the radial and the size evolution of dust in an un-split, fully implicit way.

% ..............................................................................
\subsection{Advection-diffusion Algorithm}\label{app:alg_advdif}
% ..............................................................................
To be able to also model both, the evolution of dust and gas implicitly, we constructed a scheme which solves a general form of an advection-diffusion equation,
\begin{equation}
\ddel{\N}{t} + \ddel{}{x} \left( \N \cdot u\right) -  \ddel{}{x} \left[ h \cdot D_\text{d} \cdot \ddel{}{x} \left(g\cdot \frac{\N}{h}\right) \right] = K + L \cdot \N
\label{eq:adv_diff}
\end{equation}
which can be adapted to both, Eq. \ref{eq:ssd} and Eq. \ref{eq:dustequation} by proper choice of parameters $u$, $D_\text{d}$, $g$, $h$, $K$ and $L$.

We use a flux-conserving donor-cell scheme which is implicit in $\N$.
The time derivative in Eq. \ref{eq:adv_diff}, written in a discretized way becomes
\begin{equation}
\ddel{N}{t} \hat{=} \frac{\N^{i+1}_n - \N^{i}_n}{t_{i+1}-t_{i}}
\end{equation}
where $i$ denotes time-dimension and $n$ denotes space-dimension.

The advective part is discretized as
\begin{equation}
\ddel{}{x} \left( \N \cdot u\right) = \frac{F_{n+\half}^{i+1} \cdot S_{n+\half}}{V_n} - \frac{F_{n-\half}^{i+1} \cdot S_{n-\half}}{V_n}
\label{eq:advection_num}
\end{equation}
where $F^{i+1}_{n+\half}$ and $S_{n+\half}$ are the future flux and the surface between cell $n$ and cell $n+1$ and $V_n$ is the volume of cell $n$.
The advective interface fluxes can then be written as
\begin{eqnarray}
F_{n+\half}^{i+1} &=& \N_n \cdot \max(0,u_{n+\half}) + \N_{n+1} \cdot \min(0,u_{n+\half}) \\
F_{n-\half}^{i+1} &=& \N_{n-1} \cdot \max(0,u_{n-\half}) + \N_n \cdot \min(0,u_{n-\half}) 
\end{eqnarray}
The diffusive interface flux becomes
\begin{eqnarray}
F_{d,n+\half}^{i+1} &=& D_{\text{d},n+\half} \: h_{n+\half} \: \frac{\frac{g_{n+1}}{h_{n+1}}\N_{n+1} - \frac{g_{n}}{h_{n}}\N_{n}}{x_{n+1}-x_{n}}\\
F_{d,n-\half}^{i+1} &=& D_{\text{d},n-\half} \: h_{n-\half} \: \frac{\frac{g_{n}}{h_{n}}\N_{n} - \frac{g_{n-1}}{h_{n-1}}\N_{n-1}}{x_{n}-x_{n-1}}
\end{eqnarray}
Working out these equations and separating the values of $\N$ leads to
\begin{equation}
\N_n^i = A_n \cdot \N_{n-1}^{i+1} + B_n \cdot \N_{n}^{i+1} + C_n \cdot \N_{n+1}^{i+1} + D_n
\label{eq:matrix_r}
\end{equation}
with the coefficients
\begin{equation}
\begin{array}{lll}
A_n &=& -\frac{\Delta t}{V_n} \left( \max(0,u_{n-\half}) \cdot S_{n-\half} + \frac{D_{\text{d},n-\half} \cdot h_{n-\half} \cdot S_{n-\half} \cdot g_{n-1}}{(x_n-x_{n-1}) h_{n-1}} \right)\\
B_n &=& 1- \Delta t  L_n + \frac{\Delta t}{V_n} \Bigg( \max(0,u_{n+\half}) \cdot S_{n+\half} \\
&&- \min(0,u_{n-\half})\cdot S_{n-\half}\\
&&+\frac{D_{\text{d},n+\half}\cdot h_{n+\half} \cdot S_{n+\half} \cdot g_n}{(x_{n+1}-x_n) h_n} + \frac{D_{\text{d},n-\half} \cdot h_{n-\half}\cdot S_{n-\half} \cdot g_n}{(x_n-x_{n-1}) h_n} \Bigg)\\
C_n &=& \frac{\Delta t}{V_n} \left( \min(0,u_{n+\half})\cdot S_{n+\half} - \frac{D_{\text{d},n+\half}\cdot h_{n+\half}\cdot S_{n+\half}\cdot g_{n+1}}{(x_{n+1}-x_n) h_{n+1}} \right)\\
D_n &=& -\Delta t \cdot K_n.
\end{array}
\label{eq:coefficients_r}
\end{equation}
Eq. \ref{eq:matrix_r} can now be solved by any matrix-solver, but since it is a tri-diagonal matrix, the fastest analytical way to solve it is by \emph{forward elimination/backward substitution}.

It should be noted that Eq. \ref{eq:adv_diff} is implicit only in $\N$ which means that the equations we solve are only implicit in the surface density. In the case of the viscous accretion disk, described by Eq. \ref{eq:ssd}, we face the problem that also the turbulent gas viscosity $\nug$ depends on the temperature which (in the case of viscous heating) depends on the surface density. This can cause numerical instabilities to develop.

To stabilize the code, we use a scheme which estimates the temperature in several predictor steps. The actual time step is then done with the predicted temperature.

% ..............................................................................
\subsection{Coagulation Algorithm}\label{app:alg_coag}
Discretizing Eq. \ref{eq:smolu} on a mass grid $m_i$ gives
\begin{equation}
\ddel{}{t} n_k(r,z) = \sum_{ij} M_{ijk} \,n_i(r,z)\, n_j(r,z),
\end{equation}
where the dust particle number density is
\begin{equation}
n_i(r,z) = \int_{m_{i-1/2}}^{m_{i+1/2}} n(m,r,z)\, \dx m
\end{equation}
If we assume that the coagulation and fragmentation kernels are constant in $z$ and that the vertical distribution of grains is a Gaussian with a scale height according to Eq. \ref{eq:h_dust},
\begin{equation}
n_k(r,z) = \frac{N_k(r)}{\sqrt{2\pi}h_k(r)} \cdot \exp\left(-\frac{z^2}{2h_k(r)^2} \right),
\end{equation}
we can vertically integrate the coagulation/fragmentation equation.
\begin{equation}
\begin{split}
\ddel{}{t} N_k(r)
&= \int_{-\infty}^\infty \dot n_k \: \text{d}z\\
&= \sum_{ij} M_{ijk} \: N_i(r) \: N_j(r) \: \frac{1}{2\: \pi \: h_i \: h_j} \times\\
&\qquad\times\int_{-\infty}^\infty \exp{\left(-\frac{h_i^2+h_j^2}{2h_i^2h_j^2} \cdot z^2\right)} \text{d}z\\
&= \sum_{ij}  \widetilde M_{ijk} \: N_i(r) \: N_j(r),
\end{split}
\label{eq:smolu_mdiscrete}
\end{equation}
where
\begin{equation}
\widetilde M_{ijk} = \frac{1}{\sqrt{2\:\pi\:\left(h_i^2+h_j^2\right)}}\: M_{ijk}.
\end{equation}
Discretizing Eq. \ref{eq:smolu_mdiscrete} also in radial direction and rewriting it in terms of the quantity
\begin{equation}
\N_{kl} = N_k(r_l)\cdot r_l
\end{equation}
yields
\begin{equation}
\ddel{}{t} \N_{kl} = {\displaystyle \sum_{ij}}  \frac{\widetilde M_{ijk}}{r_l} \cdot \N_{il}\cdot \N_{jl} := {S}_{kl}.
\label{eq:smolu_in_u}
\end{equation}
where the vector $\mathbf{S}=\{S_{kl}\}_{k=1,m}$ is the source function for each of the $m$ mass bins.

The numerical change of $\N_{kl}$ within a time step $\Delta t = t_i-t_{i-1}$ is $\Delta \N_{kl} = \N_{kl}^{i+1}-\N_{kl}^{i}$. The time-discretized version of Eq. \ref{eq:smolu_in_u} then becomes (omitting second order terms)
\begin{equation}
\begin{split}
\frac{\Delta \N_{kl}}{\Delta t} &= \sum_{ij}  \frac{\widetilde M_{ijk}}{r_l} \: \left(\N_{il} + \Delta \N_{il}\right) \: \left(\N_{jl} + \Delta \N_{jl}\right)\\
&=\sum_{ij} \frac{\widetilde M_{ijk}}{r_l} \: \left( \N_{il} \N_{jl} + \Delta \N_{il} \N_{jl} + \N_{il} \Delta\N_{jl} + \cancel{\Delta \N_{il} \Delta \N_{jl}}\right).
\end{split}
\end{equation}
Since the first term on the right hand side is the explicit source function, we can write
\begin{equation}
\begin{split}
\frac{\Delta \N_{kl}}{\Delta t} &= S_{kl} + \sum_{ij} \frac{\widetilde M_{ijk} + \widetilde M_{jik}}{r_l}\cdot \N_{jl} \cdot \Delta \N_{il}.
\end{split}
\end{equation}
Using the vectors $\Delta \mathbf{N}= \{\Delta \N\}_{k=1,n_m}$ and $\mathbf{N}= \{\N\}_{k=1,n_m}$, this can be rewritten in matrix notation,
\begin{equation}
\left( \frac{\mathbbm{1}}{\Delta t} -\mathbf{J} \right) \Delta \mathbf{N} = \mathbf{S}.
\label{eq:coag_matrix_equation}
\end{equation}
Were
\begin{equation}
J_{ki} = \sum_{j} \frac{\widetilde{M}_{ijk} + \widetilde{M}_{jik} }{r_l}\:\N_{jl}
\end{equation}
denotes the Jacobian of the source function and $\mathbbm{1}$ the unity matrix.
The solution for the future values can now be derived by inverting the matrix in Eq. \ref{eq:coag_matrix_equation},
\begin{equation}
\begin{split}
\mathbf{N}^{i+1} &= \mathbf{N}^{i}+\Delta \mathbf{N} \\
&= \mathbf{N}^{i}+ \left( \frac{\mathbbm{1}}{\Delta t} -\mathbf{J} \right)^{-1} \cdot \mathbf{S}\\
\end{split}
\end{equation}

% ..............................................................................
\subsection{Fully implicit scheme for radial motion and coagulation}\label{app:alg_both}

\begin{figure}[t]
  \centering
  \resizebox{0.7\hsize}{!}{\includegraphics{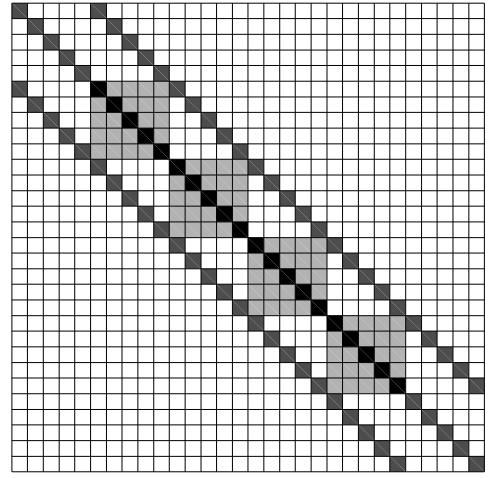}}
  \caption{Pictographic representation of the matrix on the left hand side of Eq.~\ref{eq:combined_matrix} with six radial and five mass grid points. White elements are zero, dark grey elements contain contributions from radial transport of dust ($\mathbf{J}$), light gray elements contain contributions from the coagulation/fragmentation ($\mathbf{\hat M}$), and black elements contain both contributions. The upper left and lower 5 rows represent the boundary conditions where coagulation/fragmentation is not taken into account. The matrix in the simulations would typically have a size of 15000$^2$.}
  \label{fig:matrix}
\end{figure}

To be able to solve the radial motion and the Smoluchowski equation fully implicitly, we rewrite Equation \ref{eq:matrix_r} as
\begin{equation}
\mathbf{M} \cdot \mathbf{N}^{i+1}= \mathbf{N}^i- \mathbf{D},
\label{eq:radial1}
\end{equation}
where $\mathbf{M}$ is the tri-diagonal matrix with entries $A,B$ and $C$  and source term $\mathbf{D}$ which are given by Eq. \ref{eq:coefficients_r}.
$\mathbf{M}$ is now rewritten by separating off the diagonal terms and by dividing by $\Delta t$,
\begin{equation}
\mathbf{M} = \Delta t\cdot \mathbf{\hat M} + \mathbbm{1}
\end{equation}
which brings Eq. \ref{eq:radial1} in a form similar to Eq. \ref{eq:coag_matrix_equation},
\begin{equation}
\left( \frac{\mathbbm{1}}{\Delta t} + \mathbf{\hat M} \right) \cdot \Delta \mathbf{N} = -\mathbf{\hat M} \cdot \mathbf{N}^i - \mathbf{D}/\Delta t
\end{equation}
The coagulation/fragmentation is now determined by the matrix $\mathbf{J}$ and the source vector $\mathbf{S}$ and similarly, the radial motion is determined by matrix
$\mathbf{\hat M}$ and source vector
\begin{equation}
\mathbf{R} = -\mathbf{\hat M} \cdot \mathbf{N}^i - \mathbf{D}/\Delta t.
\end{equation}
This allows us to combine both operators into one matrix equation,
\begin{equation}
\left( \frac{\mathbbm{1}}{\Delta t} + \mathbf{\hat M} -\mathbf{J} \right) \cdot \Delta \mathbf{N} = \mathbf{R} + \mathbf{S}.
\label{eq:combined_matrix}
\end{equation}
In principle, to solve for the vector $\Delta \mathbf{N}$, the matrix on the
left hand side of Eq.~\ref{eq:combined_matrix} has to be inverted. This is a
numerically challenging task since the inverse matrix in our simulations would have about 150--500 million elements. The matrix on the left hand side of Eq.~\ref{eq:combined_matrix}, however is a sparse matrix (schematically depicted in Figure~\ref{fig:matrix}).

We can therefore solve Eq.~\ref{eq:combined_matrix} by an iterative incomplete LU decomposition solver for sparse matrices provided by the Sixpack library\footnote{available from \url{www.engineers.auckland.ac.nz/~snor007}} of S. E. Norris.

\begin{figure}[t]
  \resizebox{\hsize}{!}{\includegraphics{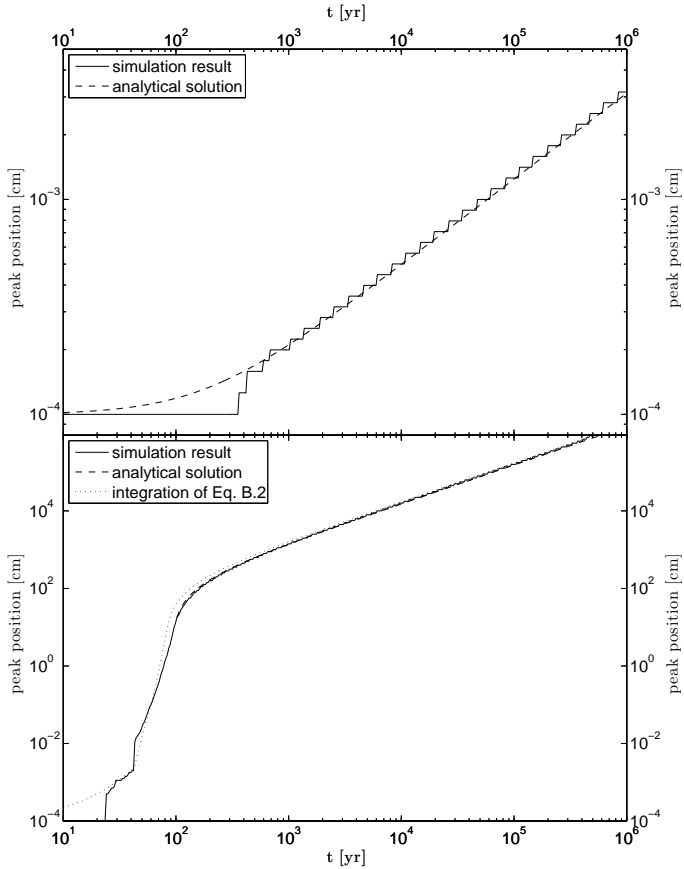}}
  \caption{Comparison between simulation result and analytical solution for growth of equal sized particles. The solid lines denote the position of the peak of the grain size distribution. In the top panel, only Brownian motion is considered as source of relative particle velocities, in the bottom panel, turbulent relative velocities are considered as well. The parameters of these simulation are $T=196$~K, $\rhos$=1.6~$\text{g/cm}^3$, $\Siggas=18$~$\text{g/cm}^2$, $\Sigdust=0.18\text{~g/cm}^2$ and $\alpha=10^{-3}$.}
  \label{fig:growthtest}
\end{figure}

% ------------------------------------------------------------------------------
\section{Test cases}\label{app:tests}
To check if the numerical implementation presented above accurately solves Eq. \ref{eq:combined_matrix}, we compare results of the simulation to some analytical solutions:
The growth rates of particles can be approximated if we assume the grain size distribution to be a delta function and the sticking probability to be unity. The increase of mass per collision is then given by the mass of the particle, $m$ divided by the time between two collisions, $\tau$, which can be written as

\begin{equation}
\frac{\Del m}{\Del t} = \frac{m}{\tau}.
\end{equation}
Using $\Del m={4\pi} \rhos a^2 \Del a$ and $\tau = m/(\rhodust \;\sigma \; \Delta u)$, we derive
\begin{equation}
\frac{\Del a}{\Del t} = \frac{\rhodust}{\rhos} \; \Delta u,
\label{eq:simple_growth}
\end{equation}
where $\Delta u$ is the relative velocity, $\rhodust$ the dust density, $\rhos$ the solid density of the dust particles and $\sigma = 4\pi a^2$ the cross section of the collision.\balance

\begin{figure}[t]
  \resizebox{\hsize}{!}{\includegraphics{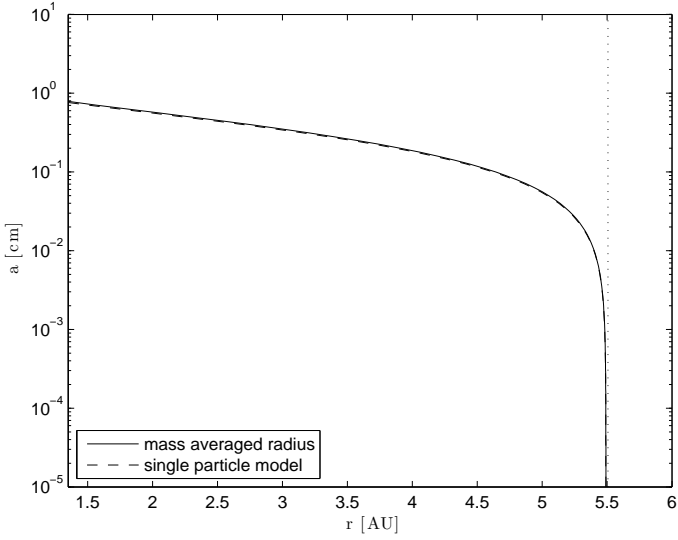}}
  \caption{Test case for only radial drift, without coagulation. The solid line denotes the mass-averaged position of the radial distribution of grains for each grain size after $10^3$~years. The dashed line is the expected solution for a single particle, the dotted line denotes the initial radius of the particles.}
  \label{fig:motiontest}
\end{figure}

With this formula, we can estimate the growth time scales if the relative velocities of the dust grains is given. For equal sized particles and Brownian motion, we get
\begin{equation}
\Delta u_\text{BM} = \sqrt{\frac{16 \; \kb T}{\pi \; m}},
\end{equation}
for turbulent motion, the relative velocities are \citep[see][]{Ormel:2007p801}
\begin{equation}
\Delta u_\text{TM} \approx \left\{ \begin{array}{ll}
\cs\sqrt{2\;\alpha \; \St}      &   \text{for }\St \ll 1\\
\cs \sqrt{\frac{2 \alpha}{\St}} &   \text{for }\St \gg 1\\
\end{array}\right..
\end{equation}
Integrating Eq. \ref{eq:simple_growth} gives
\begin{equation}
a(t) = \left( \frac{5}{2} \frac{\rho_d}{\rho_s \: \pi} \sqrt{\frac{12 \kb T}{\rhos}} \left( t - t_0\right) + a_0^{\frac{5}{2}} \right)^{\frac{2}{5}}
\label{eq:BMgrowth}
\end{equation}
for Brownian motion and
\begin{equation}
a(t) \approx \left\{\begin{array}{ll}
a_0\cdot\exp\left(\frac{\Siggas}{\Sigdust} \frac{\pi}{\sqrt{2}} \cdot(t-t_0)\right) &   \text{for }\St \lesssim 1\\
\frac{\Sigdust \Ok}{2 \rhos \sqrt{\pi}} \cdot (t - t_0) + a_0       &   \text{for }\St > 1\\
\end{array}\right.
\end{equation}
for relative velocities due to turbulent motion.

A comparison between analytical solution and simulation result for Brownian motion growth is shown in the top panel of Figure \ref{fig:growthtest}. It can be seen that the position of the peak of the grain size distribution (solid line in Figure \ref{fig:growthtest}) follows the analytical result of Eq. \ref{eq:BMgrowth}.

A similar comparison for the case of relative velocities due to turbulent motion is not as straight-forward since both the turbulent relative velocities and the dust scale height (Eq. \ref{eq:h_dust}) are sub-divided into several regimes. We therefore integrated Eq. \ref{eq:simple_growth} numerically for the case of Brownian motion and turbulent relative velocities; the results are shown in the bottom panel of Figure \ref{fig:growthtest}. As before, we see that -- after the initial condition is overcome -- the simulation result follows closely the mono-disperse approximation. For grains larger than $\St=1$, the analytical solution is also plotted in the bottom panel of Fig. \ref{fig:growthtest}, almost coinciding with the simulation results.

The radial motion of dust particle was tested in a similar fashion: starting from a grain distribution at a radius of 5.5~AU, we let the particles drift (taking the gas drag and the radial drift into account, see Eq. \ref{eq:u_r_dust}) without coagulation. We compare the results to results of a numerical integration of the equation of motion for a single particle. The results are shown in Fig. \ref{fig:motiontest}.

We find that the size distribution behaves as expected: small particles are well coupled to the gas, they almost retain their initial position since the radial motion due to gas drag is in the order of 0.01~AU. Larger particles, having a larger Stokes number drift towards the star on shorter timescales. Particles of a few centimeters (corresponding to \St=1) are already lost after about 700~years.

The mass in all test cases was found to be conserved on the order of $10^{-11}$\% of the initial value.
% ==============================================================================

\end{document}